\documentclass[useAMS, letterpaper, usenatbib, fleqn, twocolumn]{mn2e} 
\usepackage{times}
\usepackage{graphicx}
\usepackage{amsfonts}
\usepackage{amsmath}
\usepackage{amssymb}
\usepackage{rotating} 
\usepackage{setspace}
\usepackage{xspace}
\usepackage{url}
\usepackage{gensymb}
\usepackage{enumerate}
\usepackage[ table ]{ xcolor }
\usepackage{pinlabel}
\DeclareMathAlphabet{\mathbf}{OML}{cmm}{b}{it}
\DeclareMathAlphabet{\mathbfsf}{OT1}{cmss}{bx}{n}

\newcommand{\Msol}{M_{\odot}}

\newcommand\ion[2]{#1$\,${\scshape{#2}}}%
\newcommand\hi{\ion{H}{i}\xspace}

\def\ctwo{\textsc{C}$^2$\textsc{-ray}\xspace}
\def\cthree{\textsc{CubeP}$^3$\textsc{M}\xspace}

\newcommand\bq{\begin{equation}}
\newcommand\eq{\end{equation}}
\newcommand\bqs{\begin{equation*}}
\newcommand\eqs{\end{equation*}}

\newcommand\bqa{\begin{eqnarray}}
\newcommand\eqa{\end{eqnarray}}
\newcommand\bqas{\begin{eqnarray*}}
\newcommand\eqas{\end{eqnarray*}}

\newcommand{\be}{\begin{equation}}
\newcommand{\e}{\end{equation}}

\newcommand{\bear}{\begin{eqnarray}}
\newcommand{\ear}{\end{eqnarray}}

\def\D2{\Delta^2}
\def\De2L{{\Delta_{\rm LC}^2}}
\def\De2C{{\Delta_{\rm CC}^2}}
\def\xh1{{x_{\rm HI}}}
\def\k{{\bf k}}
\def\kpl{k_\parallel}

\def\mpc{{\rm Mpc}}
\def\mpci{{\rm Mpc^{-1}}}
\newcommand{\deltahi}{\delta_{\rho_{\mathrm{HI}}}}

\def\beq{\begin{equation}}
\def\eeq{\end{equation}}

\title[Light cone effect on the reionization 21-cm signal]
{Light cone effect on the reionization 21-cm signal II: Evolution, anisotropies and observational implications}

\author[Datta et al.]{Kanan K. Datta $^{1,2}$ 
\thanks{e--mail: kanan@ncra.tifr.res.in},
Hannes Jensen$^{2}$, 
Suman Majumdar$^{2}$,
Garrelt Mellema$^{2}$,
\thanks{e--mail: garrelt@astro.su.se},
Ilian T. Iliev$^{3}$,
\newauthor Yi Mao$^{4}$,
Paul R. Shapiro$^{5}$
and Kyungjin Ahn$^{6}$
\\
$^1$National Centre For Radio Astrophysics, Post Bag 3, Ganeshkhind, Pune 411 007, India\\
$^2$Department of Astronomy \& Oskar Klein Centre, AlbaNova, Stockholm University, SE-106 91 Stockholm, Sweden \\
$^3$Astronomy Centre, Department of Physics \& Astronomy, Pevensey II Building, 
University of Sussex, Falmer, Brighton BN1 9QH\\
$^4$Institut Lagrange de Paris, Institut d’Astrophysique de Paris, CNRS, UPMC Uiversite Paris 06, UMR7095, 98 bis, boulevard Arago, F-75014 Paris, France\\
$^5$Department of Astronomy and Texas Cosmology Center, University of Texas, Austin, Texas 78712, USA \\
$^6$Department of Earth Sciences, Chosun University, Gwangju 501-759, Korea\\
 }

\date{\today}
\pubyear{2012} \volume{001} \pagerange{1}
\begin{document}
\maketitle
\begin{abstract}
Measurements of the HI 21-cm power spectra from the reionization epoch will be influenced by the evolution of the signal along the line-of-sight direction of any observed volume. We use numerical as well as semi-numerical simulations of reionization in a cubic volume of 607 Mpc across to study this so-called light cone effect on the HI 21-cm power spectrum. We find that the light cone effect has the largest impact at two different stages of reionization: one when reionization is $\sim 20\%$ and other when it is $\sim 80\%$ completed. We find a factor of $\sim 4$ amplification of the power spectrum at the largest scale available in our simulations. We do not find any significant anisotropy in the 21-cm power spectrum due to the light cone effect. We argue that for the power spectrum to become anisotropic, the light cone effect would have to make the ionized bubbles significantly elongated or compressed along the line-of-sight, which would require extreme reionization scenarios. We also calculate the two-point correlation functions parallel and perpendicular to the line-of-sight and find them to differ. Finally, we calculate an optimum frequency bandwidth below which the light cone effect can be neglected when extracting power spectra from observations. We find that if one is willing to accept a $10 \%$ error due to the light cone effect, the optimum frequency bandwidth for $k= 0.056 \, \rm{Mpc}^{-1}$ is $\sim 7.5$ MHz. For $k = 0.15$ and $0.41  \, \rm{Mpc}^{-1}$ the optimum bandwidth is $\sim 11$ and $\sim 16$ MHz respectively.

\end{abstract}
\section{Introduction}
A period of major changes in the Universe took place between $\sim0.3$ to $\sim1$ billion years after the Big Bang at the time when the first
galaxies and supermassive black holes were forming. These first
structures produced ionizing radiation which escaped from the parent
environment to spread through the intergalactic medium (IGM) and
ionize the neutral hydrogen (\hi) in it. This period is known as the
Epoch of Reionization (EoR). Our knowledge about this epoch in the
evolution of the Universe is currently very limited as even the most
powerful telescopes can only offer a glimpse of objects from this
time. Observations of the redshifted 21-cm signal from \hi in the IGM
are expected to unveil how the reionization process progressed and
will thus provide us with invaluable information about this epoch
\citep[see e.g.][for reviews]{2012RPPh...75h6901P, 2013ExA....36..235M}.

A great deal of effort during recent years, both in theory and in
observations, has put ``reionization with 21-cm radiation'' at the
forefront of the modern astronomy \citep{2010ARA&A..48..127M}. The
first generation of low-frequency radio telescopes such as Low Frequency
Array \footnote{http://www.astro.rug.nl/eor} \citep{2013A&A...556A...2V}, Giant Metrewave Radio Telescope \footnote{http://gmrt.ncra.tifr.res.in/} \citep{2008AIPC.1035...75P},
Murchison Widefield Array \footnote{http://www.mwatelescope.org} \citep{2013PASA...30....7T} , Precision Array for Probing
the Epoch of Reionization \footnote{http://eor.berkeley.edu/} \citep{2010AJ....139.1468P}  have started providing us with results such as preliminary measurements of the foregrounds at EoR frequencies \citep{2008MNRAS.385.2166A,
  2009MNRAS.399..181P, 2009A&A...500..965B, 2012MNRAS.426.3295G,
  2013A&A...550A.136Y, 2013ApJ...771..105B, 2013ApJ...776..108J} as
well as limits on reionization \citep{2010Natur.468..796B,
  2013MNRAS.433..639P, 2013arXiv1304.4991P}.

These telescopes may be able to detect large individual ionized
regions embedded in \hi
\citep{2007MNRAS.382..809D,2008MNRAS.386.1683G,2012MNRAS.424..762D,2012MNRAS.426.3178M,2013ApJ...767...68M} or provide crude maps of
large \hi regions \citep{2012MNRAS.425.2964Z,2013MNRAS.429..165C}.
However, they will primarily try to detect the EoR \hi 21-cm signal
statistically, using quantities such as the power spectrum
\citep{2004ApJ...608..622Z, 2004ApJ...615....7M, 2010MNRAS.405.2492H},
visibility correlations \citep{2005MNRAS.356.1519B,
  2008MNRAS.385.2166A}, variance \citep{2006MNRAS.372..679M,2008MNRAS.389.1319J,
  2011JCAP...04..038B,2014arXiv1401.4172P} and 
skewness \citep{2009MNRAS.393.1449H} of the
\hi 21-cm brightness temperature fluctuations. Understanding these
statistical quantities and their connection to the physics of reionization
is crucial in planning the observational strategies, analysing
and validating the observations and extracting the reionization
characteristics from them. Substantial efforts on the theoretical side
have been undertaken towards this. For example, analytic models of
reionization which are very fast but less accurate have been developed
\citep{2004ApJ...613....1F}. Despite many challenges, there has been
significant progress in estimating the large-scale \hi 21-cm signal
from the entire EoR through numerical \citep{2006MNRAS.371.1057I,
  2006MNRAS.372..679M, 2007MNRAS.376L..34M, 2007MNRAS.377.1043M,
  2008ApJ...681..756S, 2009A&A...495..389B} and semi-numerical
simulations
\citep{2007ApJ...654...12Z,2007ApJ...669..663M,2008MNRAS.386.1683G,2008ApJ...689....1S,
  2009MNRAS.393...32T,2009MNRAS.394..960C,2013ApJ...776...81B}.

Numerical simulations of the EoR which calculate the ionization state of the IGM by solving the proper radiative transfer equation at each cell in the simulation box can be highly accurate. Such simulations can conserve the total photon numbers emitted by all sources, introduce complex photon production histories for the sources and can take into account the effect of recombination self-consistently. However, they do require substantial
computational resources in order to achieve this \citep{2014MNRAS.439..725I}. On the other hand, semi-numerical simulations, which simply compare the total number of ionizing photons available and the number of baryons to be ionized in order to calculate the ionization state, require much less resources. Although approximate in nature, recent studies show that at least for simplified source models, semi-numerical simulations can produce reasonable results with up to $\sim 10\%$ error in the power spectrum estimates \citep[see e.g.][for a recent study]{2014arXiv1403.0941M}.  The availability 
of numerical and semi-numerical simulations allow us to explore various 
source  models and study important issues such as the light cone effect, 
redshift space
distortions, sinks, very bright QSOs, different feedback mechanisms,
etc., on the observable statistical quantities mentioned above. The
simulated signal can also be used to determine the detectability of
various observables, explore different foreground subtraction methods
and test for other systematics.

In this paper we study in detail the impact of the so-called \emph{light
cone effect} (LC effect) on the EoR \hi 21-cm signal. Since the signal originates
from a line transition, different cosmological epochs correspond to
different wavelengths, characterized by their cosmological redshift
$z$ through
\begin{equation}
  \lambda_{\rm obs}=\lambda_{\rm emitted}(1+z)\,.
\end{equation}
This means that an observational data set containing a range of
wavelengths corresponds to a time period in which the signal may have
evolved. This is commonly known as the light cone effect. The
radio telescope data will be in the form of three-dimensional image
cubes of images over a wide range of frequencies. The analysis of
these 3D data sets should generally take into account this LC effect.

In \citet{2012MNRAS.424.1877D} (hereafter Paper I) we presented the
first numerical investigation of the LC effect on the spherically-
averaged \hi 21-cm power spectrum. We used the results of large volume
numerical simulations ($163 \, \mpc$ (comoving) on each side) and
studied the effect at four different stages of reionization. We found
that the effect mostly `averages out', but can lead to a change of $\sim50\%$
in the power spectrum at scales around $k\sim 0.1 \,
\mpc^{-1}$. We also attempted to study the LC anisotropy in the signal
but the simulation volume was too small to quantify this properly. In
the analysis we also incorporated redshift space distortions (due
to the gas peculiar velocity) and found them to have a negligible
impact on the LC effect (see section 5 and figure 14 in Paper I). 

\citet{plante2013} used a larger volume\footnote{The size
  of the full box is $2/h \, {\rm Gpc}$, but they use
  sub-volumes of line-of-sight  extent up to $500 \, {\rm Mpc}$
  for the LC analysis.} and focused on similar issues. They reported
significant anisotropies in the full signal but also noted that radio
interferometric experiments, which only measure fluctuations in the
signal and therefore exclude certain Fourier modes, will observe the
signal to be isotropic. 

Another way to characterize the LC effect is to compare the two-point
correlation function of the signal along and perpendicular to the
line-of-sight. \citet{barkana2006} adopted this approach in the first,
analytic study of LC anisotropies in the 21-cm signal.
\citet{2014arXiv1401.1807Z} followed the same approach, using the results of
large scale reionization simulations of size $\approx 571 \, \mpc$. 
They extended their
investigations to the pre-reionization epochs where the spin
temperature of \hi is lower or comparable to the Cosmic Microwave
Background (CMB) temperature and the signal may appear in absorption
against the CMB. Both these works report anisotropies due to the LC
effect although the properties of these anisotropies differ between
them.

In this paper we use the largest radiative
transfer simulation of the EoR to date to study the LC effect further
and on larger scales than we did in Paper I.  Although the radiative
transfer simulation is more accurate, the associated computational costs make it difficult to explore a range of parameters. To
explore the LC effect for other possible reionization models and study the robustness
of some of our findings, we use a semi-numerical reionization simulation of the same volume. Additionally we employ some toy models to understand our
findings. We focus on the following issues:
\begin{enumerate}
\item Evolution of the LC effect. At what stages and scales of
reionization is the effect important? How do redshift space distortions
and the LC effect interact?
\item Anisotropy in the 21-cm power spectrum. Does the LC effect
  introduce any anisotropy? 
\item Two-point correlation analysis of the 21-cm signal. How does the LC effect
  show up and does it introduce anisotropy?
\item Impact on analysis strategy. What frequency bandwidths can be used
  to minimize the LC effect?
\end{enumerate}

The outline of the paper is as follows. Section 2 briefly describes
our reionization simulations, numerical and semi-numerical, and how we
produce mock 21-cm datacubes including the LC effect and redshift space
distortions. We introduce the \hi 21-cm power spectrum and its
anisotropies in Section 3. We present our results on the impact of the
LC effect on the 21-cm power spectrum, both the spherically averaged
version and the anisotropy, in Section 4.  Section 5 presents the
results of simple analytic and toy models which help in understanding
our LC anisotropy results and provide insight into which conditions
introduce LC anisotropies in the power spectrum. Results on the two-point 
correlation function are described in Section 6. Section 7
discusses the impact of the LC effect on the power spectrum derived
from observations. Here we also provide an optimal frequency bandwidth below
which the complications due to the LC effect are mostly
avoided. Finally, we summarize the results and conclude in Section 8.

We assume a flat $\Lambda$CDM model with parameters $\Omega_m=0.27$, $\Omega_b=0.044$, $h=0.7$, $n=0.96$, $\sigma_b=0.8$, consistent with the nine year {\it Wilkinson Microwave Anisotropy Probe results} \citep{2013ApJS..208...19H} and Planck \citep{2013arXiv1303.5076P}.

\section{Simulating the reionization 21-cm signal}
\subsection{$N$-body and radiative transfer simulations}
The basis of our reionization simulations is a (607 Mpc)$^3$ $N$-body
simulation performed with the \cthree code and post-processed with the
\ctwo \, code (see \citet{2014MNRAS.439..725I} for details). \cthree \,
\citep{2013MNRAS.436..540H} is a cosmological $N$-body code based on
\textsc{PMFast} \citep{2005NewA...10..393M}. It calculates
gravitational forces on a particle-particle basis for short distances
and using a grid for long distances. Here, we used 5488$^3$ particles,
each with a mass of $5\times 10^7 \Msol$, and a grid with 10976$^3$
cells  which gives a spatial resolution of $\sim 5.5 \,$ kpc.  We note that in a P$^3$M $N$-body code the resolution is determined by the gravitational force softening instead of the cell size. Outputs from this simulation have been recorded at intervals of
$11.5$ Myrs in the
redshift range $z=6$ to $30$. This yields a total $76$
simulation boxes of dark matter distribution. For each \cthree output,  we 
use a halo finder to locate dark matter halos and measure their masses.  The  halo finder is based on the spherical over-density method and a group of more than 20 particles with over-density
$\Delta >178$  with respect to the mean dark matter density is considered to be halo. This provides us with a list of dark matter halos with masses down to $\sim 10^9 \; \Msol$ together with their positions. 

It is believed that dark matter halos with masses as low as $\sim 10^8 \; \Msol$ would be able to form stars through atomic cooling and thus contribute to reionization. To resolve such low-mass halos, we would need to use $10$ times more particles compared to what we use here, which is computationally unfeasible. Instead, we use a sub-grid recipe calibrated through a higher mass resolution but
smaller volume simulation (volume (163 Mpc)$^3$, $3072^3$ particles) to introduce halos down to $10^8 \;
\Msol$. The results of this simulation give us the total mass in halos of masses between $10^8$ and $10^9 \;\Msol$ as a function of redshift and density in
volumes of size $1.2$ Mpc, this being the size of cells in our radiative
transfer grid. Our method reproduces the mean dark matter halo mass function for the entire box but it neglects the scatter in the local over-density-halo number relation which is observed numerically. We then combine the local total mass for the missing halos with the dark matter halo list obtained directly by the halo finder method to create a full list of ionizing sources.

Using the halo lists and density field outputs---down-sampled to
504$^3$ cells or a resolution of 1.2 Mpc---we then simulated the reionization of the IGM using
\ctwo \citep{2006NewA...11..374M}. \ctwo works by casting rays from
ionizing sources and iteratively solving for the evolution of the
ionized fraction of hydrogen in each point on the grid. The dark
matter halos from the $N$-body simulations were used as ionizing
sources, which each halo with mass $M_{\mathrm{h}}$ assumed to have an
ionizing flux:
\begin{equation}
	\dot{N}_{\gamma} = g_{\gamma} \frac{M_{\mathrm{h}}
          \Omega_\mathrm{b}}{(10\;\mathrm{Myr}) \Omega_\mathrm{m}
          m_{\mathrm{p}}},
	\label{eq:ionizing_flux}
\end{equation}
where $m_{\mathrm{p}}$ is the proton mass and $g_{\gamma}$ is a source
efficiency coefficient, effectively incorporating the star formation
efficiency, the initial mass function and the fraction of ionizing
photons escaping into the IGM. Here, we use:
\begin{align}
	g_{\gamma} = 
	\begin{cases} 
		1.7 & \text{for } M_{\mathrm{h}} \geq 10^9 \Msol \\
		7.1 & \text{for } M_{\mathrm{h}} < 10^9 \Msol.
\end{cases}
\label{eq:g_gamma}
\end{align}
motivated by the fact that low-mass halos should have a higher escape
fraction and more top-heavy initial mass function. These low-mass
halos are turned off completely when the local ionized fraction
exceeds $10\%$ since they lack the gravitational well to keep
accreting material in a highly ionized environment
\citep{2007MNRAS.376..534I}.

\begin{figure*}
	\begin{center}
		\includegraphics{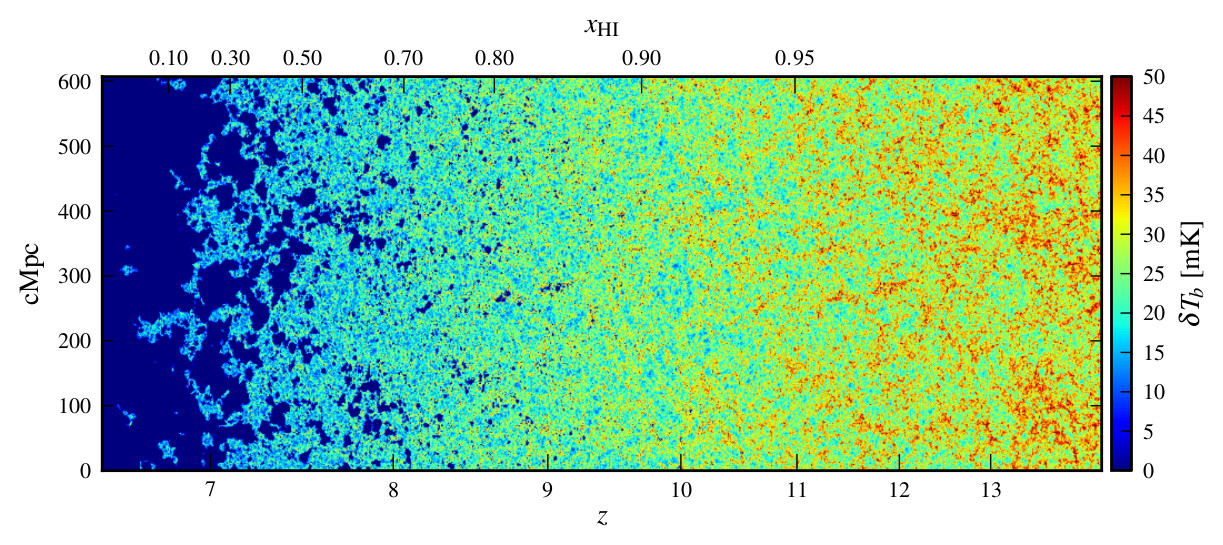}
	\end{center}
	\caption{A slice through the 21-cm brightness temperature
          lightcone volume from the numerical simulation.}
	\label{fig:lightcone_slice}
\end{figure*}

\subsection{Lightcone volumes}
From the steps described above we obtain a series of simulation
volumes of the dark matter distribution and the ionization state of
the IGM, each at a constant redshift (``coeval'' cubes). We then
calculate the \hi 21-cm brightness temperature, $\delta T_b(\bold{r},
z)$, in these volumes as:
\begin{equation}
	\delta T_b(\mathbf{r}, z) = \widehat{\delta T_b}(z)[1 + \deltahi(\mathbf{r})],
	\label{eq:brightnesstemp_short}
\end{equation}
where $\widehat{\delta T_b}(z)$ is the global mean 21-cm brightness
temperature at redshift $z$ and $\deltahi(\mathbf{r})$ is the the
over density of neutral gas in some point $\mathbf{r}$. This equation
is only valid once the spin temperature of the IGM is much higher than
the CMB temperature, which should be true in all but the earliest
stages of reionization.

Next, we need to transform these into lightcone volumes, i.e.\ volumes
where the evolution state of the IGM changes along the
line-of-sight. We do this by stepping through redshifts and for each
redshift $z$, find two coeval cubes whose redshifts bracket $z$ and
interpolate between these two cubes. The exact procedure is described
in detail in Paper I.

After this step, we end up with a volume consisting of cells of
constant comoving size, but where the evolutionary state of the IGM
along the line-of-sight matches what an observer would actually see. A
slice through the \hi 21-cm brightness temperature from the LC volume
is shown in Fig.\ \ref{fig:lightcone_slice}. Note that, although our LC
volume extends to around $1821 \, \mpc$ along the line-of-sight axis, we only
use at most $607\, \mpc$ for our analysis. Since the size of our individual
simulation cubes are $(607 \, \mpc)^3$, any statistics for larger distances
will be affected by periodicity effects.

\subsection{Redshift space distortions}
We also wish to study the LC effect together with the redshift space distortions introduced by gas peculiar velocities. Since the redshift of a parcel of gas is determined not only by its cosmological redshift but also by the line-of-sight component of its peculiar velocity, the signal that one would reconstruct by translating redshifts to line-of-sight positions is not the same as the real-space signal.  Matter tends to flow towards high-density regions, and away from low-density voids, which causes the 21-cm signal in redshift space to  become anisotropic. The signal also has a higher contrast than in real space. See \citet{2005MNRAS.356.1519B} and  \cite{2012MNRAS.422..926M} for detailed reviews of the theory of redshift space distortions.

We incorporate this effect using the MM-RRM scheme described in detail in \cite{2012MNRAS.422..926M}. In short, this method works by moving the boundaries of each cell according to the gas velocity at the position at the boundary. A cell boundary at a real-space position $\mathbf{r}$ is moved to a redshift space position $\mathbf{s}$:

\begin{equation}
	\mathbf{s} = \mathbf{r} + \frac{1 +
          z_{\mathrm{obs}}}{H(z_{\mathrm{obs}})} v_{\parallel} (t,
        \mathbf{r}) \hat{r},
	\label{eq:reddist}
\end{equation}
where $z_{\mathrm{obs}}$ is the observed redshift of the cell and
$v_{\parallel}$ is the line-of-sight component of the gas peculiar velocity at
this position.  Typically, gas velocities are on the order of a few hundred km/s, resulting in cell boundary displacements that can reach up to 2 or even 3 cMpc in the most dense regions. Since we interpolate the velocity over the cell, any velocity gradient across the cell will cause the cell to be stretched or compressed in redshift space and the 21-cm signal to either decrease or increase, respectively. After this step, we re-grid the redshift space data to the original cell size. This entire procedure is carried out after constructing the LC data volumes, since applying the effects in the other order would lead to unphysical results at the edges of the coeval data volumes.

\subsection{Semi-numerical Simulations}

We have just one realization of reionization based on the radiative
transfer simulation described earlier in this section. However, there
is considerable uncertainty in the nature and luminosity of the
sources of reionization. For example, reionization may be considerably
earlier and faster than in our numerical simulation. It is anticipated
that such a rapid reionization may enhance the light cone anisotropy
in the signal \citep{barkana2006}. However, it is computationally
expensive to rerun our radiative transfer simulation to generate many
different scenarios.  We side step this issue by using a
semi-numerical method to simulate such an early and rapid reionization
and test whether that enhances the light cone anisotropy in the
observables of the 21-cm signal.

This semi-numerical simulation employs the method described in
\citet{2009MNRAS.394..960C}. This method is inspired by the
excursion-set formalism \citep{2004ApJ...613....1F} and closely
follows the methodology described in \citet{2007ApJ...669..663M}, with
some differences. Like all other reionization models, here we also
assume that the ionizing photons were produced in dark matter
halos. Due to the lack of knowledge about the properties of the
sources of these photons we assume that the total number of ionizing
photons produced in a halo of mass $M_{\rm h}$ is
\begin{equation}
N_{\gamma}(M_{\rm h}) = N_{\rm ion} \frac{M_{\rm h}}{m_{\rm p}} 
\end{equation}
where $N_{\rm ion}$ is a dimensionless constant. We use the same
$N$-body mass distribution and halo catalogues as that of the
radiative transfer simulation described earlier. To simulate an
ionization map at a specific redshift, we calculate the average
ionizing photon number density $\langle n_{\gamma} ({\bf x})\rangle_R$
and \hi atom number density $\langle n_{\rm H} \rangle_R$ within a
spherical region of radius $R$ around a point ${\bf x}$ and compare
them. This averaging and comparison is done for a range of smoothing
scales, starting from the cell size ($R_{\rm cell}$) up to a certain
maximum radius $R_{\rm max}$, which is decided by the photon mean free
path at that redshift. We consider the point ${\bf x}$ to be ionized
only if the condition
\begin{equation}
  \langle n_{\gamma} ({\bf x})\rangle_R \ge \langle n_{\rm H}
  \rangle_R (1+ {\bar N}_{\rm rec})
\end{equation}
(eq. [7] of \citet{2009MNRAS.394..960C}) is satisfied for any
smoothing radius $R$. The factor ${\bar N}_{\rm rec}$ is the average
number of recombinations per hydrogen atom in the IGM. Note that
various other unknown parameters e.g. the star forming efficiency
within the halos, the number of photons per unit stellar mass, the
photon escape fraction, the helium weight fraction, as well as the
factor $(1 + {\bar N}_{\rm rec})$ are absorbed within the
definition of $N_{\rm ion}$. Given the mass
and location of the halos at a specific redshift we tune the parameter
$N_{\rm ion}$ to obtain a desired neutral fraction.  Here we do not
consider a density dependent recombination scenario and the effect of
self-shielding, which can be added in these simulations
\citep{2009MNRAS.394..960C}. We set an ionization fraction equal to
$\langle n_{\gamma} ({\bf x})\rangle_{R_{\rm cell}}/\langle n_{\rm H}
\rangle_{R_{\rm cell}}$ to the grid points where the above condition
is not fulfilled. Finally, we tune the value of $N_{\rm ion}$ at each
redshift to achieve a desired reionization history, i.e.\ $ \xh1 $
vs. $z$.

\begin{figure}
	\begin{center}
		\includegraphics[width=\columnwidth]{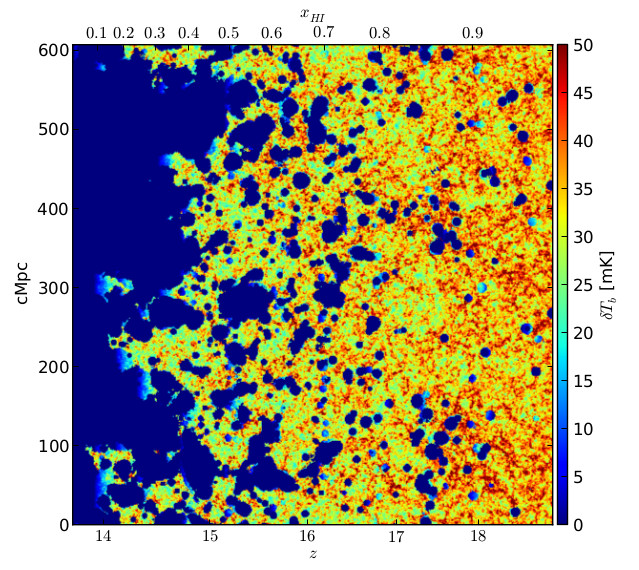}
	\end{center}
	\caption{A slice through the 21-cm brightness temperature
          lightcone volume from the semi-numerical simulation.}
	\label{fig:lightcone_slice_sem_num}
\end{figure}

For our study we consider an extreme reionization scenario which
starts at $z\sim20$ and ends at $z\sim13.5$.  A light cone slice
through the 21-cm brightness temperature from the semi-numerical
simulation is shown in
Fig.\ \ref{fig:lightcone_slice_sem_num}. In this scenario,
$10\%$, $50\%$ and $90 \%$ (mass averaged) reionization occurs at
redshifts $17.85$, $15.2$ and $14$ respectively. The time period
between $10\%$ and $90\%$ reionization is just $\sim 88$ Myr---much faster than in our full numerical simulation for which the similar
time scale is $\sim 400\,$Myr. This rapid scenario is inconsistent
with the latest results from the WMAP satellite \citep{2013ApJS..208...19H}.
However, it is illustrative of scenarios in which a few relatively bright sources drive reionization and has a reionization history which closely matches the
``Pop III`` case in \citet{barkana2006}.


\begin{figure*}
\includegraphics[width=1.0\textwidth]{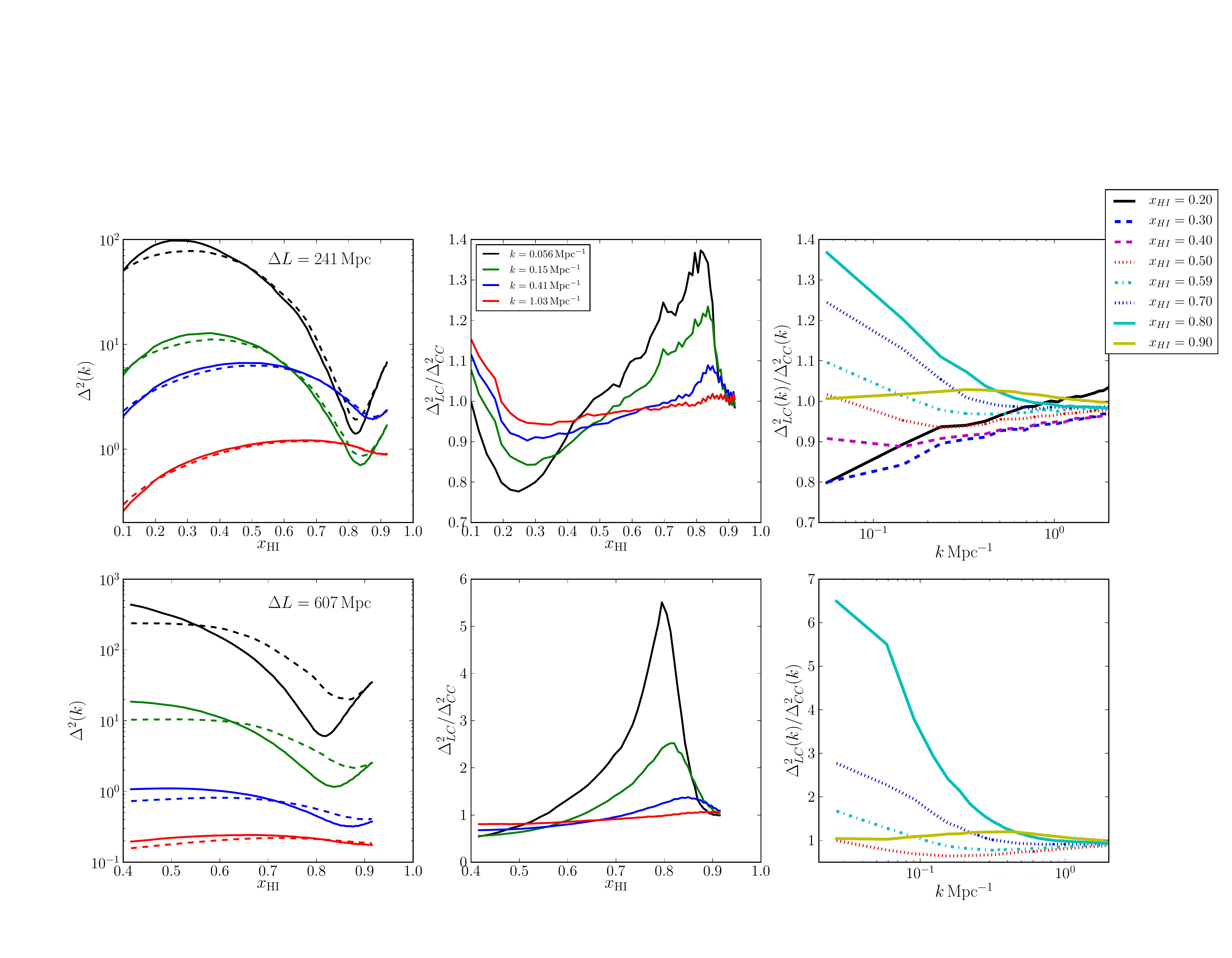}
\caption{The LC effect on the spherically-averaged power spectrum $\D2 (k)$ of the \hi 21-cm brightness temperature fluctuations. The upper panel considers a sub-volume of the line-of-sight extent $\Delta L=241 \, \mpc$  from a full simulation box and the lower panel shows results for full box of size $607 \, \mpc$. The solid and dashed lines in the upper left panel represent the evolution of the spherically-averaged power spectrum without and with the LC effect respectively as a function of the neutral fraction $\xh1$ for different $k$ values ($k=0.056$, $0.15$, $0.41$, $1.03$ $\mpci$ from the top to bottom). Note that the power spectrum amplitudes for different $k$-modes are fixed arbitrarily for clarity in presentation. Here the redshift space distortions have not been included. The middle and right panels show the ratio between power spectra with and without the LC effect as function of the neutral fraction $\xh1$ and $k$-mode respectively. }
\label{fig:ratio}
\end{figure*}
\begin{figure*}
\includegraphics[width=1.0\textwidth]{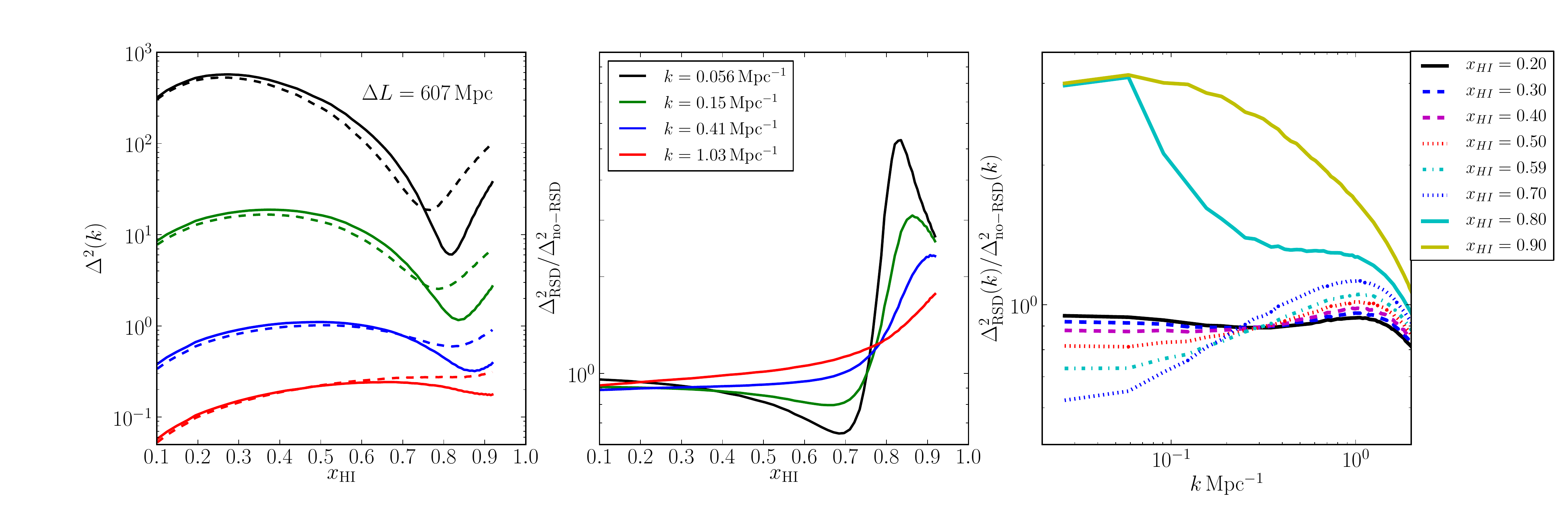}
\caption{Same as the lower panel of the Figure \ref{fig:ratio}, except here we include the redshift space distortion effect but not the LC effect. The solid and dashed lines in the left panel correspond to the spherically averaged power spectrum without and with the redshift distortion effect respectively.}
\label{fig:ratio-pv-nopv}
\end{figure*}

\begin{figure*}
\includegraphics[width=1.0\textwidth]{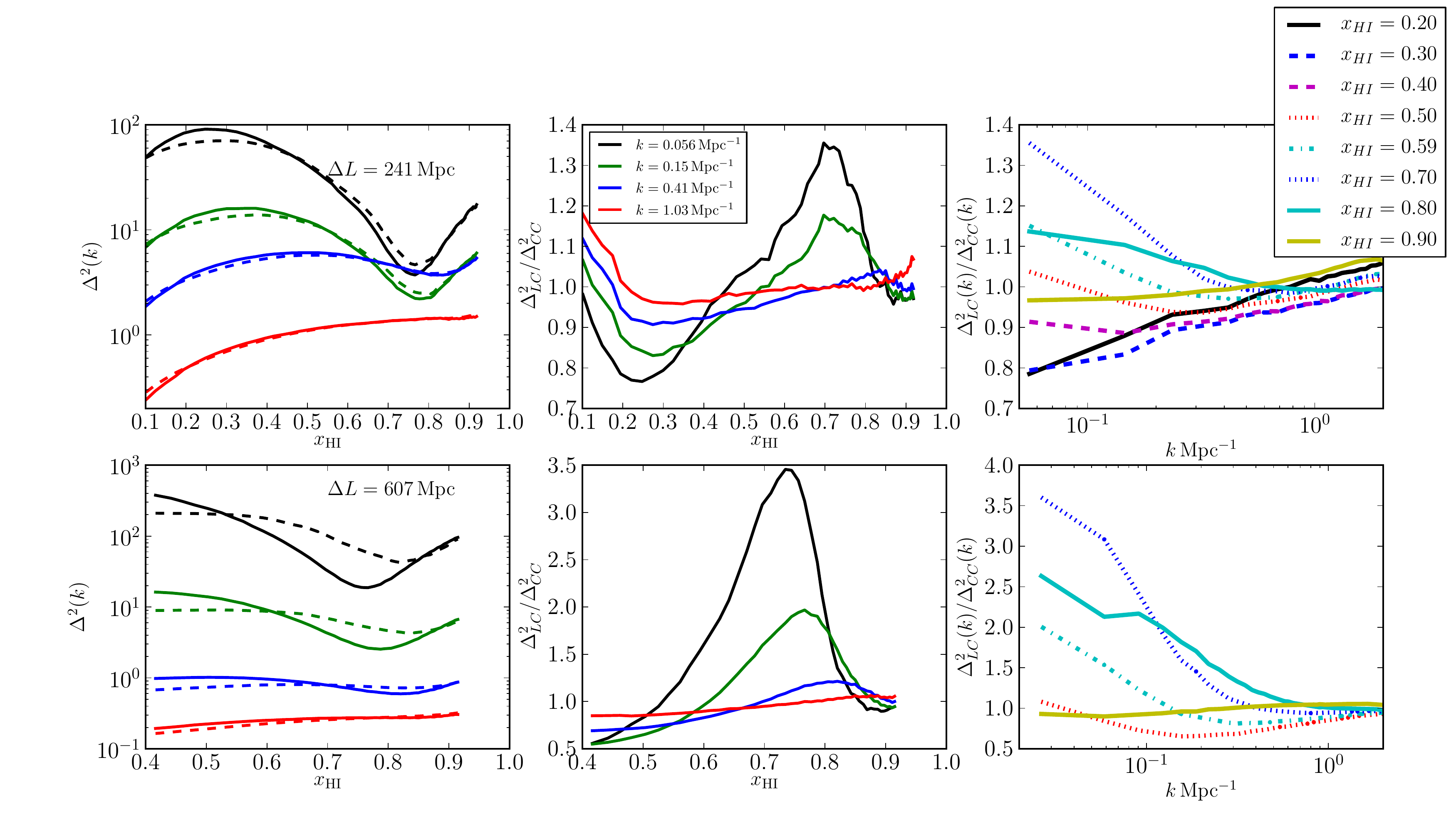}
\caption{Same as Figure \ref{fig:ratio}, but we include redshift space distortions while estimating power spectra with and without the LC effect.}
\label{fig:ratio-pv}
\end{figure*}

\section{The 21-cm power spectrum and its anisotropies}
Although at large scales our Universe is isotropic, there are several effects that introduce differences between the \hi 21-cm power spectrum $P({\bf k})$ measured along and perpendicular to the line-of-sight. To quantify this anisotropy in the power spectrum 
the 21-cm power spectrum is often written as  $\D2 (k,\mu)$, where $\mu=\kpl/k$ and $\kpl$ is the line-of-sight component of the Fourier mode $\k$. The dimensionless power spectrum $\D2$ which is defined as $\D2=k^3P(k)/2 \pi^2$ essentially measures the variance of fluctuations in the \hi brightness maps at scales correspond to mode $k$. The parameter $\mu$  which is also written as $\cos \theta$ runs from $-1$ to $1$, $\theta$ is the angle between the line-of-sight and the mode $\k$. $|\mu|=1$ and $0$ corresponds to the signal along and perpendicular to the line-of-sight respectively. If the power spectrum is instead integrated over all $\mu$ values we refer to it as the spherically-averaged \hi 21-cm power spectrum, $\D2(k)$.

One effect that makes the 21 cm power spectrum anisotropic is due to the gas peculiar velocities along the line of sight, which displace the 21 cm signal from its cosmological redshift. This is known as redshift space distortions and these carry unique information about the underlying matter density and the cosmic reionization \citep{2005ApJ...624L..65B,2012MNRAS.422..926M,2013PhRvL.110o1301S, 2013MNRAS.435..460J, 2013MNRAS.434.1978M}. The other effect is the Alcock-Paczynski effect which occurs when the use of inaccurate cosmological parameters introduce a mismatch in the calculation of physical scales along and perpendicular to the line of sight \citep{1979Natur.281..358A, 2005MNRAS.364..743N, 2005MNRAS.363..251A, 2006MNRAS.372..259B}. 

In the linear regime  the anisotropic \hi 21-cm power spectrum arising from the above two effects can be written as \citep{2006MNRAS.372..259B}:
\begin{equation}
\D2 (k,\mu)=\D2_0(k)+\D2_{\mu^2}(k)\mu^2+\D2_{\mu^4}(k)\mu^4+\D2_{\mu^6}(k)\mu^6.
\end{equation} 
$\D2_0(k)$ is the power spectrum that would be observed if there were no redshift space distortions and no Alcock-Paczynski effect in the signal. $\D2_{\mu^2}(k)$  is the  cross-correlation power spectrum between the ionization and matter density fluctuations. $\D2_{\mu^4}(k)$ is the power spectrum of total matter density fluctuations multiplied by the mean \hi brightness temperature squared. The Alcock-Paczynsky effect introduces a $\mu^6$ dependence in the power spectrum. The LC effect, being a line of sight effect, could in principle introduce additional anisotropies. In the subsequent sections we study the anisotropies due to the LC effect. However, before we do so we first consider how the LC effect affects the spherically-averaged \hi 21-cm power spectrum.

\section{Results}
\subsection{Impact on the spherically-averaged power spectrum}
\label{sect:sph_avg_ps}
Fig. \ref{fig:ratio}  shows the LC effect on the spherically-averaged  \hi 21-cm power spectrum. We do not include $k_{\perp}=0$ modes throughout our work as this mode is not measurable by interferometric experiments. In this figure, we only present results from our numerical simulation. The redshift space distortion effect has not been included in this figure. The upper panel is for a sub-volume with line-of-sight extent $\Delta L=241 \, \mpc$ from the full simulation box and the lower panel shows results for the full box of size $607 \, \mpc$. The choice of $\Delta L=241 \, \mpc$ allows us to probe the neutral fraction $\xh1$ as low as $0.1$. Smaller $\Delta L$ will allow us to probe the LC effect for smaller $\xh1$. As we see later the LC effect is negligible for $\xh1<0.1$, and so we do not consider sub-volumes of $\Delta L<241 \, \mpc$.  The dashed and solid lines in the upper left panel show the evolution of the spherically-averaged power spectrum with and without the LC effect respectively as a function of the mass-averaged neutral fraction $\xh1$ for different $k$ modes ($k=0.056$, $0.15$, $0.41$, $1.03$ $\mpci$ from the top to bottom). Note that the power spectrum amplitudes for different $k$ modes are fixed arbitrarily to minimize overlap between the curves. 

First considering the $241 \, \mpc$ results, we find that at large scales ($k \lesssim 0.4 \, \mpci$) the power spectrum without the LC effect initially goes down  when  $\xh1$ decreases from $1$. There is a minimum in the power spectrum without the LC effect around $x_{\rm HI}\sim 0.8$--$0.9$ for all $k$ modes we consider. During these early stages of reionization mainly high density peaks get ionized and therefore large scale \hi fluctuations are suppressed. This dip in the spherically averaged power spectrum has been seen in other works \citep{2006MNRAS.371.1057I, 2013MNRAS.435..460J, 2013MNRAS.434.1978M}. 

Beyond $x_{\rm HI}\sim 0.8$ the power at larger scales starts to grow rapidly as ionized bubbles grow substantially and provide fluctuations to the \hi field at large scales. At the late stages of reionization the power in the non-LC results again starts to drop due to rapid decline in $x_{\rm HI}$ and finally becomes zero at the end of the EoR. We find that the evolution in the spherically-averaged power spectrum as a function of $\xh1$ (or redshift) is most dramatic at large scales and is more gradual at smaller scales. For example, the spherically-averaged power spectrum for $k=0.056 \, {\rm \mpci}$ changes by around two orders of magnitude during the time period when $x_{\rm HI}$ changes from $0.8$ to $0.2$. At the smallest scale shown the changes are less than one order of magnitude.

The spherically-averaged power spectrum with the LC effect included (dashed lines in the upper left panel) is higher compared to the power spectrum without the LC effect during the initial stages ($x_{\rm HI}\sim 0.8 $ to $0.9$) and lower at the late stages of reionization for all $k$ modes we consider. This can be clearly seen in the upper-middle panel where we plot the ratio $\frac{\D2_{\rm LC}}{\D2_{\rm CC}}(k)$ between the spherically-averaged power spectra with and without the LC effect. The ratio peaks around $x_{\rm HI}\sim 0.8$--$0.9$ and dips around $x_{\rm HI}\sim 0.2$--$0.3$. For $k=0.056 \, {\rm \mpci}$ the ratio goes up to $\sim 1.4$ at $x_{\rm HI}\sim 0.8$  and dips down to  $\sim 0.8$ at $x_{\rm HI} \sim 0.25$.  The ratio gradually approaches $1$ at small scales.  We note that during these two different stages of reionization the evolution of the spherically-averaged power spectrum without the LC effect is highly non-linear and rapid. Around $x_{\rm HI}\sim 0.8 $ to $0.9$ there is a dip in the power spectrum and it grows rapidly on either side of this dip. The large-scale power spectrum peaks at around $x_{\rm HI}\sim 0.2 $ to $0.4$ and decreases rapidly on both sides of this peak. 

This non-linear evolution of the power spectrum makes the LC effect very strong. On the other hand, the power spectrum around $\xh1 \sim 0.4-0.5$ evolves more linearly with $\xh1$ and hence the ratio is very close to $1$, i.e.\ the LC effect is small. Therefore we conclude that the LC effect for a given $k$ mode is mainly determined by the non-linear evolution of the power spectrum with the neutral fraction $\xh1$. This confirms our conclusion in Paper I---that the LC effect will be determined by the terms $\frac{d^{2n} \D2 (k,z)}{dr^{2n}}$ where $r$ is the comoving distance to redshift $z$ and $n>=1$.

The upper right panel of Fig.\ \ref{fig:ratio} shows the ratio between the spherically-averaged power spectra with and without the LC effect as a function of $k$-mode for different neutral fractions $\xh1$. We find that initially when $x_{\rm HI}\gtrsim 0.9$ (solid yellow line) the ratio is close to $1$, with slightly higher values at a scale of $0.4 \, \rm {Mpc}^{-1}$.   We also find that the ratio gradually increases at larger scales during the first half and decreases in the second half of the EoR. 

The bottom panels are the same as the upper panel except the line-of-sight length of the volume is $\Delta L=607 \, {\rm Mpc}$ in this case.  Here we can not study the LC effect for the small neutral fraction $\xh1<0.4$ as LC volumes of line-of-sight width of  $\Delta L=607 \, {\rm Mpc}$ centered around lower neutral fraction and redshift would extend beyond the redshift range of our simulations. The main difference between the 241 and $607 \, {\rm Mpc}$ results is that in the latter the ratio between the  spherically-averaged power spectra with and without the LC effect is much higher but otherwise the results are qualitatively similar. For example for $k \sim 0.056\, \rm {Mpc}^{-1}$ the ratio goes up to $\sim 5.5$ around $\xh1 \sim 0.8$. A larger impact of the LC effect is expected as the larger volume encompasses a much larger redshift range and therefore evolution of the 21 cm signal. 

These results show that the power spectra of 21-cm data-sets of very large frequency widths will be easier to measure, at least from the early stages of reionization, but also that we need to take the LC effect into account when analyzing and interpreting such data-sets. We will discuss the optimum line-of-sight extent for analyzing EoR \hi 21-cm data in Section \ref{sec:optimum-bw}.

\subsection{Impact of  redshift space distortions on the spherically-averaged power spectrum}
In order to compare the impact of the LC effect and the redshift space distortion effect we start by presenting in Fig. \ref{fig:ratio-pv-nopv} the spherically-averaged \hi 21-cm power spectrum in redshift space but without the LC effect. The $k$ modes are the same as in  Figure \ref{fig:ratio} and the line-of-sight extent of the underlying simulation is $607 \, \mpc$. Interestingly, we find some similarities between the ways the redshift space distortions and the LC effect affect the spherically-averaged  \hi 21-cm power spectrum.  For both the power spectrum is affected more strongly at larger scales. The effect is most dramatic at the initial stages ($\xh1 \sim 0.8$) of the EoR where the ratio between the spherically-averaged  \hi 21-cm power spectra with and without the redshift space distortion effect reaches up to a factor of $6$ for $k=0.056 \, \mpci$. However at small scales and during the later stages of the EoR this ratio comes close to $1$. In contrast, the redshift space distortion effect is very small at the late stages of the EoR ($\xh1 \lesssim 0.3$) whereas the LC effect has a significant effect at both the early and late stages. The redshift space distortion effect does not depend on the line-of-sight length of the volume analyzed whereas the LC effect for obvious reasons does.

Fig. \ref{fig:ratio-pv} shows the same results as Fig. \ref{fig:ratio} but including the redshift space distortion effect while calculating the spherically-averaged power spectra with and without the LC effect. Although results are qualitatively same as in Fig. \ref{fig:ratio} we see that the peak in the ratio moves to lower $\xh1 \sim 0.7$ and the peak height is also lower compared to the case without redshift space distortions.  We found similar results in our previous study (Paper I).

The enhancement of the spherically-averaged power spectrum due to the LC effect around $\xh1 \sim 0.8-0.9$ is not seen in \citet{plante2013}. We find that this enhancement is due to the fact that the power spectrum without the LC effect becomes very small around that stage of reionization (see Figs. \ref{fig:ratio}, \ref{fig:ratio-pv}) and therefore the ratio gets boosted. We believe that this dip in the power spectrum without the LC effect shows up  due to the high mass resolution of our simulations.  We use $5488^3$ DM particles, a total of $10976^3$ cells in a comoving volume of $(425/h \, \mpc)^3$ , whereas the above mentioned work uses $2048^3$ DM particles in a comoving volume of $2/h, {\rm Gpc}$. This gives us $\sim 2000$ times better particle mass resolution. Because of the coarser resolution in \citet{plante2013}, the lower mass halos which form early in the high density peaks are not included and thus they will lack the early suppression of these high density peaks.

\subsection{Anisotropies in the power spectrum}
\label{subsec:anisotropies}
\begin{figure*}
\includegraphics[width=1.0\textwidth]{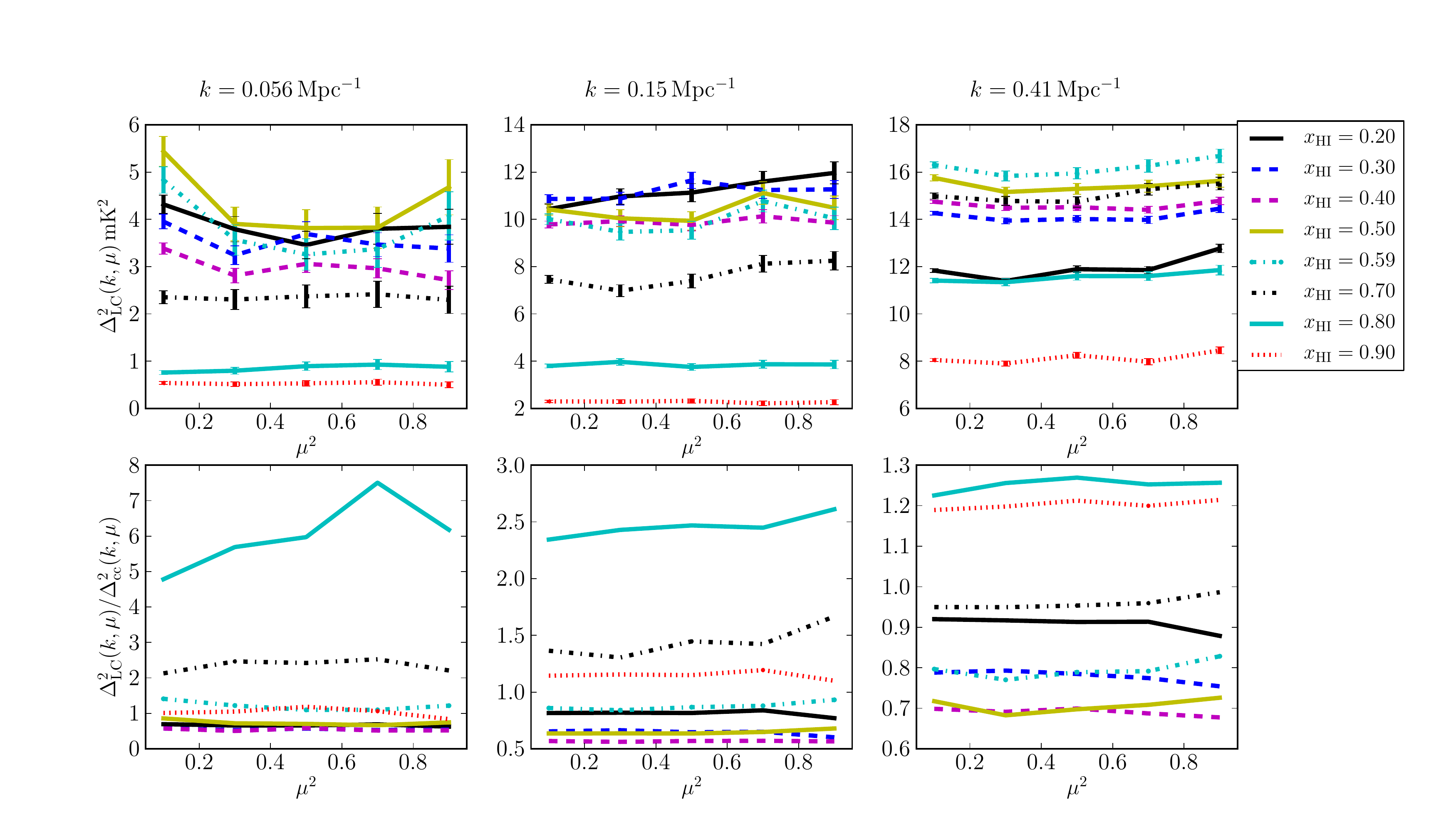}
\caption{The power spectrum $\Delta^2(k,\mu)$ (upper panels) with the LC effect and the ratio between power spectra with and without the LC effect (lower panels) as a function  $\mu^2$  for a given $k$-mode. Different lines in each panel represent different neutral fraction ranging from $x_{\rm HI} = 0.2 $ to $0.9$. The $k$-mode is fixed for each panel.}
\label{fig:aniso-c2ray1}
\end{figure*}

\begin{figure*}
\includegraphics[width=1.0\textwidth]{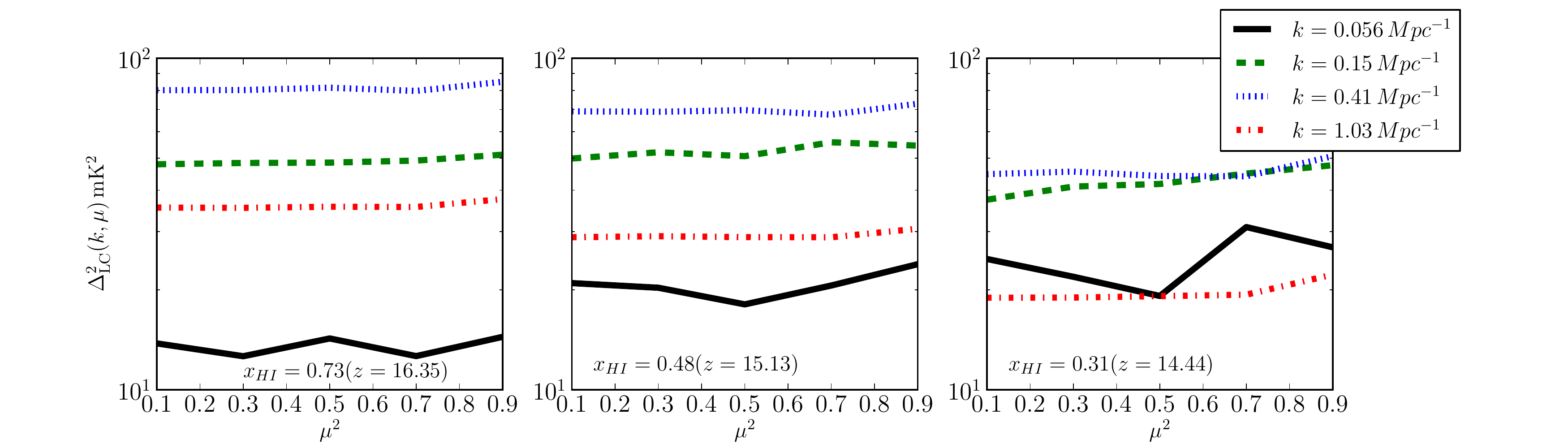}
\caption{Power spectra as a function of $\mu^2$ for different $k$-modes in the semi-numerical simulations.}
\label{fig:aniso-semi-num}
\end{figure*}

We next consider the anisotropy in the observed \hi 21-cm signal. Fig. \ref{fig:aniso-c2ray1} shows the \hi 21-cm power spectrum $\Delta^2(k,\mu)$ with the LC effect (upper panels) and the ratio between the power spectra with and without the LC effect (lower panels) as a function of $\mu^2$  for different $k$ modes. As above we do not include $k_{\perp}=0$ modes here as these are unobservable with interferometers. Since we would like to investigate the anisotropy due to the LC effect only, we do not include the redshift space distortions here. Different lines in each panel represent different neutral fractions ranging from $x_{\rm HI} \sim 0.2 $ to $0.9$ for a fixed $k$-mode. The $k$-modes are $0.056$, $0.15$ and $0.41 \, \mpci$ from the left to right panel respectively. Note that the line-of-sight extents for different lines in each panel are different. When extracting sub-volumes with smaller line-of-sight extents than the full extent of the simulation we always make sure that the mean neutral fraction $\xh1$ for the coeval box and the central two-dimensional slice of the corresponding LC sub-volume are the same. The comoving distances from the centre to the two edges along the line-of-sight of the LC sub-volume are also kept equal. The line-of-sight extent is the same for the coeval and light cone sub-volumes for a fixed $k$-mode and neutral fraction $x_{\rm HI}$.  Since our focus is on the anisotropies, the amplitudes are not important here.  We include $1\sigma$ error bars due to sample variance. These have been calculated from $\sqrt{2} \Delta^2(k,\mu)/\sqrt{N}$, where $N$ is the number of $k$-modes in the range $k$, $k+dk$ and $\mu$, $\mu+d\mu$. 

The black solid line shows results for the neutral fraction $\xh1 \sim 0.2$. The non-smooth behavior of the line arises because of the small number of available modes  for it. The LC volume around $x_{\rm HI} \sim 0.2$ (redshift $z=6.94$) is only $337 \,{\rm Mpc}$ wide along the line-of-sight (compared to $607 \,{\rm Mpc}$ of full box) for the same reason as we used the $241 \,{\rm Mpc}$ extent above in Sect.~\ref{sect:sph_avg_ps}: reionization in our simulation finishes around redshift  $z=6.48$. Therefore to have a LC box centered around redshift $z=6.94$ and one end at redshift  $z=6.48$ we can have a volume with at most $337 \,{\rm Mpc}$ along the line-of-sight. The small number of $k$-modes increases the error bars for this line.  

Inspection of the top row of Fig.~\ref{fig:aniso-c2ray1} appears to reveal some variations of $\Delta^2(k)$ with $\mu$, although comparison of different stages of reionization and $k$-modes does not suggest any systematic effects. In fact, when one compares these results to the ones without the LC effect it becomes apparent that the latter show similar features. In principle the power spectra without the LC effect should not show any such $\mu$-dependence. This means that the apparent systematic change is not due to the LC effect but arises by chance. The way we calculate the error bars is valid if the underlying field is Gaussian. The reionization 21-cm signal is highly non-Gaussian, especially at late stages of the EoR, and therefore error bars may have been under-estimated. 

This point is further illustrated by the lower panels of Fig.\ \ref{fig:aniso-c2ray1} where we plot the ratio between the power spectra with and without the LC effect. We observe that the ratio does not change significantly with $\mu^2$. We also do not observe any systematic changes with $\mu^2$. From all this we conclude that if the LC effect introduces any dependency on $\mu^2$ in the 21~cm power spectrum, it is so weak that it is buried in the sample variance. 

To investigate whether a more rapid reionization scenario driven by fewer sources produces a stronger anisotropy (as claimed by \citealt{barkana2006}), we show in Fig.\ \ref{fig:aniso-semi-num} the anisotropy results for our semi-numerical simulation. We show results around three central neutral fractions: $\xh1=0.77, 0.48$ and $0.31$ (from the left to right panel). Even though reionization in the semi-numerical simulation is $\sim 4.5$ times faster than in numerical simulation and takes place much earlier (in the redshift range $\sim 14-19$), we still do not find any significant anisotropies in the \hi 21-cm power spectra. We therefore conclude that the LC effect either does not make the EoR \hi 21-cm power spectra anisotropic i.e. a function of $\mu^2$ or this dependence is so weak that it is buried in the sample variance. This is consistent with the recent work by \citet{plante2013} who also do not find any significant anisotropy due to the LC effect if $k_{\perp}=0$ modes are excluded. 

We would like to reiterate here that the LC effect does have an overall effect on the power spectra as we observed in Fig. \ref{fig:ratio}, i.e. the power spectrum is either enhanced or suppressed due to the effect but that enhancement/suppression is the same for all $\mu$ for a given $k$ and therefore does not introduce any observable $\mu$-dependence. 

\section{Toy models}
\label{sec:toymodel}
\begin{figure}
\includegraphics[width=0.4\textwidth]{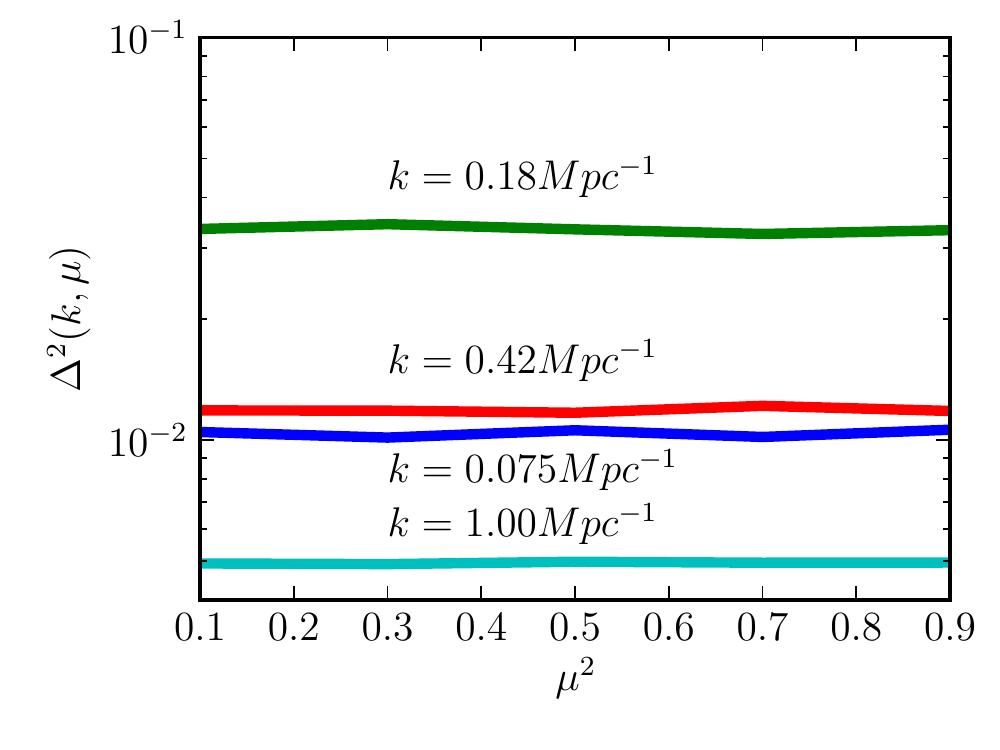}
\caption{Power spectra as a function of $\mu^2$ for different $k$-modes in the the toy model described in Section \ref{sec:toymodel} }
\label{fig:toymodel}
\end{figure}

To better understand the lack of anisotropy due to the LC effect in the simulation results, we here use some simple toy models.

Let us consider $N$ randomly placed spherical, non-overlapping ionized bubbles embedded in a uniform \hi distribution in a cube. For simplicity we assume that the signal from ionized and neutral regions are zero and one respectively. The Fourier transform for a single spherical ionized bubble of radius $R$ at position ${\bf r}$ is
\be
\label{equation1}
\delta ({\bf k}) = A (kR) \exp ^{i{\bf k}.{\bf r}}, 
\e
where $A(kR)=4/3\pi R^3\,  W(kR)$. The spherical top-hat window function is defined as $W(kR)=\frac{3}{k^3R^3}(\sin(kR)-kR \cos(kR))$. The Fourier transform for $N$ randomly placed, non-overlapping, spherical bubbles of arbitrary sizes can  be simply written as:
\bear
    \label{deltak2}
    \delta ({\bf k}) &=& \sum_{j=1}^{N} A_j \exp ^{i{\bf k}.{\bf r_j}} \nonumber \\
&=& \sum_{j=1}^{N} A_j  \left[\cos({\bf k}.{\bf r_j})+ i \,\sin({\bf k}.{\bf r_j}) \right], \nonumber\\
&&
\ear
where $A_j$ denotes $A(kR_j)$. 
The power spectrum $P({\bf k})$ for such a scenario can be written as:
\bear
    \label{pk1}
P({\bf k})&=& \langle \delta ({\bf k}) \delta^{*} ({\bf k}) \rangle \nonumber \\
&=& \left(\sum_{j=1}^{N} A_j  \cos({\bf k}.{\bf r_j}) \right)^2+ \left (\sum_{j=1}^{N} A_j \,\sin({\bf k}.{\bf r_j}) \right )^2 \nonumber \\
&=& \sum\limits_{j=1}^N A_j^2 +2\sum\limits_{j=1, l=j+1}^{N-1,N} A_jA_l\cos({\bf k}.{\bf {\Delta r_{jl}}}) \nonumber \\
\ear
where $\bf {\Delta r_{jl}}=\bf {r_{j}}-{r_{l}}$.

Now for the simplistic case where all bubbles are same of radius $R$, the above equation reduces to:
\be
\label{pk2}
P({\bf k})=NA^2+2 A^2\sum\limits_{j=1, l=j+1}^{N-1,N} \cos({\bf k}.{\bf {\Delta r_{jl}}})
\e
Since the bubbles are randomly placed, both ${\bf r_i}$ and ${\bf r_{j}}$ are random, therefore the difference vector ${\bf {\Delta r_{jl}}}$ is random too. The phase term ${\bf k}.{\bf {\Delta r_{jl}}}$ can assume any value between $-\pi$ to $\pi$ (assuming that the ${\bf r_i}$ vector is zero at the center of the observed volume). This makes the second term in the above equation zero. Therefore the above equation reduces to
\be
\label{pk3}
P({\bf k})=NA^2.
\e
Since $A$ is only a function of $|{\bf k}|$, the power spectrum is isotropic, which is expected for the above case.

Now we consider a case where the bubble size changes systematically along the line-of-sight but remains the same for a fixed line-of-sight location, i.e.\ a fixed redshift $z$ (we refer to the right panel of figure 9 in Paper I). This is motivated by the fact that the inclusion of the LC effect makes the bubbles appear smaller/larger in the far/near side of an observed volume (see Fig. \ref{fig:lightcone_slice} in this paper or figure 3 in Paper I). Note, however, that the individual bubbles are still spherical. In our simulations, we do not see any significant systematic elongation or compression in bubble shape due to the LC effect. This is expected as elongation or compression in bubble shapes occurs only when ionization fronts propagate relativistically \citep{2005ApJ...623..683Y,2005ApJ...634..715W,2006ApJ...648..922S,2008ApJ...673....1S,2011MNRAS.413.1409M,2012MNRAS.426.3178M}. 

We focus on bubbles only at two different line-of-sight locations, namely at $r_{\parallel_1}$ and $r_{\parallel_2}$.  In principle there will be many bubbles at those two line-of-sight locations. Averaging over all those will make the quantity $A_j A_l \cos({\bf k}.{\bf {\Delta r_{jl}}})$ in Eq. \ref{pk1} zero. Accordingly, all `cosine' terms for all possible line-of-sight combinations will be zero and therefore the above equation can be reduced to:
\be
P({\bf k})= \sum\limits_{j=1}^N A_j^2
\e
Like the previous case, the power spectrum here also is only dependent on $|{\bf k}|$ i.,e, the power spectrum is isotropic.

We tested the above analytic result using simulations. In a simulation box we put bubbles which are spherical, non-overlapping and randomly placed with a radius that changes with line-of-sight location. We assume that the bubble size changes linearly with line-of-sight location as $R = R_0 + R_0(n_\mathrm{los}/N-1/2)$ where $N=256$ is the total number of grid points along the line-of-sight and the line-of-sight index $n_\mathrm{los}$ varies from $0$ to $255$. $R_0$ is assumed to be $10\,\mpc$, so the bubble size varies linearly with the los index from $5 \, \mpc$ in the far side to $15 \, \mpc$ in the front side. The box size is $(256 \, \mpc)^3$. 

We simulate $500$ independent realizations where the distribution of bubble centers changes in each simulation. Fig. \ref{fig:toymodel} shows $\D2(k,\mu)$ as a function of $\mu^2$ for different $k$-modes. It shows the ensemble average of all of the independent realizations. We find that  the power spectrum is not changing with $\mu^2$, i.e. the power spectrum is isotropic  and therefore supports our analytic predictions.  We have also considered other models for the bubble size as a function of line-of-sight location, such as an exponential change where bubble size changes more drastically. We do not observe any anisotropy in any of those models. If the ionized bubbles were either prolate or oblate shaped of same size and oriented in the same direction, the power spectrum would become anisotropic as the Fourier transform of a prolate or oblate spheroid is not spherically symmetric. However, as noted above, this is only true in the case of relativistically propagating ionization fronts.

  Based on the above results we conclude that the systematic change in the bubble size as a function of line-of-sight while individual bubble remains spherical is not enough to make the power spectrum anisotropic. Individual bubbles should be either systematically elongated or compressed to make the power spectrum anisotropic. In standard reionization by galaxies (like the ones we consider here) the ionized bubbles only appear to be smaller (in the far end) or larger (in near end) without any systematic elongation or compression. This explains the  non-observance of significant anisotropy in the simulated power spectrum.

The above toy model considers only non overlapping bubbles. Due to the clustering of ionizing sources ionized bubbles overlaps with each other and therefore becomes highly non spherical with some elongation or compression along the line-of-sight. Since the overlap of bubbles has no preferred direction, the signal will remain statistically isotropic. Therefore, our conclusions are also valid for the case where ionized bubbles overlap with each other.

\section{The two-point correlation function}
\label{Sec:coor-func}

\begin{figure*}
\includegraphics[width=0.5\textwidth]{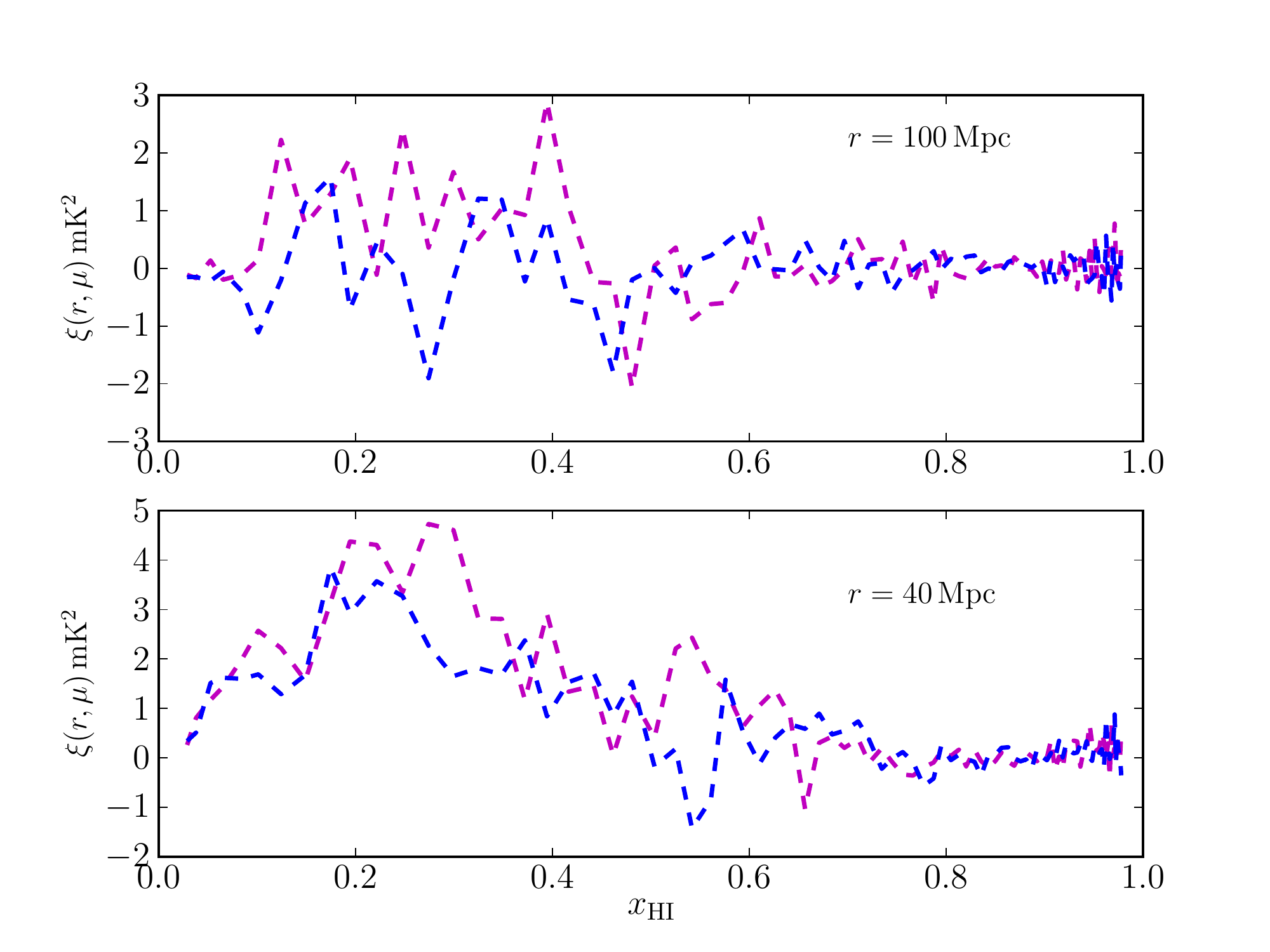}
\caption{The two-point correlation function as a function of neutral fraction $\xh1$ for a given correlation length $r$ for the numerical simulation. We use the light cone volume for this calculation. The magenta and blue lines show results for $\mu_r=1$ and $\mu_r=0$ respectively. For  $\mu_r=1$ we correlate points taken from two different slices along the line of sight which are respectively $r$ and $r/2$ distance apart from each other and from the centre corresponding to neutral fraction $x_{\rm HI}$. For $\mu_r=0$ we correlate points taken from the 2D slice corresponding to  neutral fraction $x_{\rm HI}$.}
\label{fig:corr-func0}
\end{figure*}

\begin{figure*}
\includegraphics[width=0.5\textwidth]{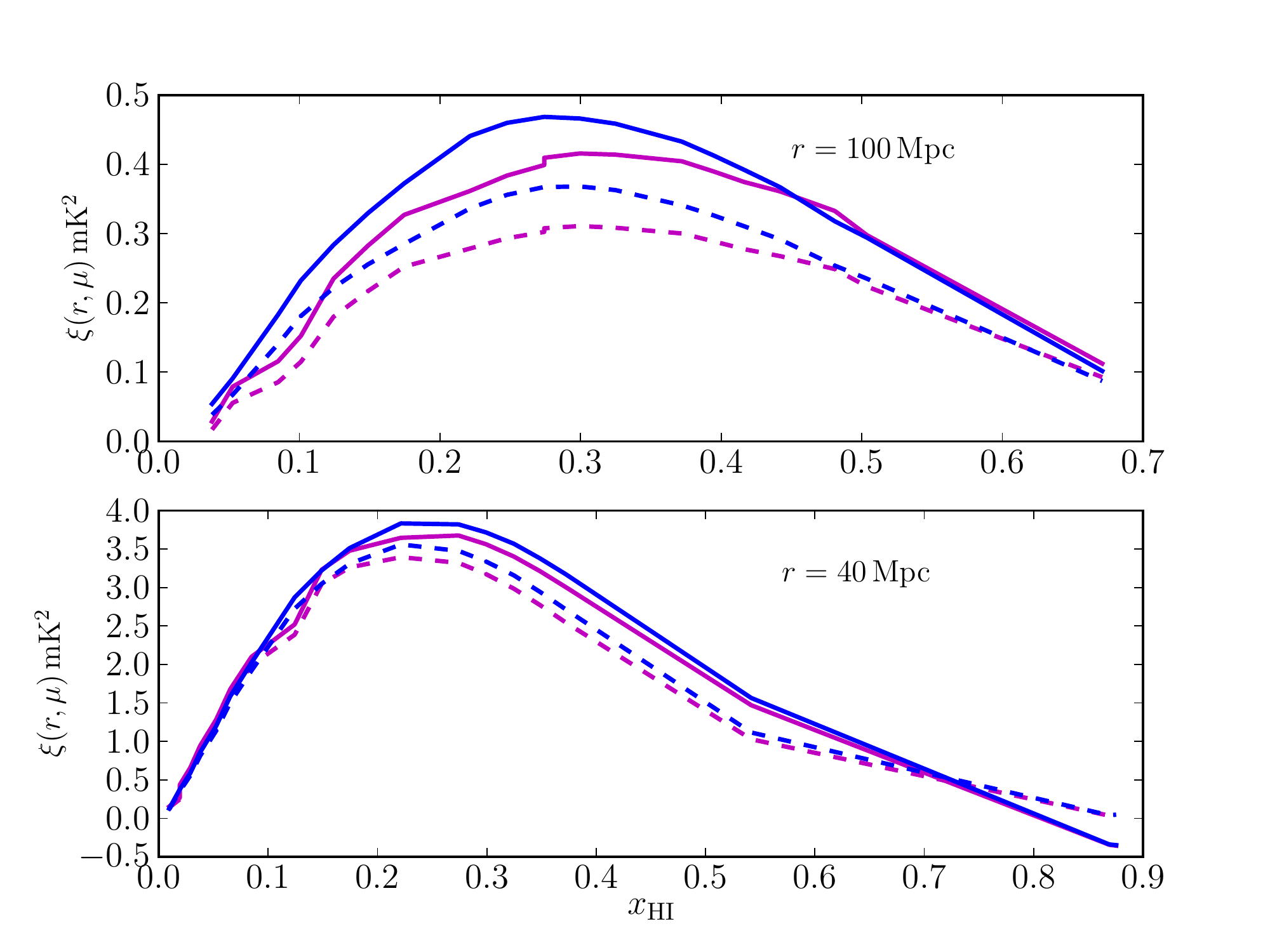}
\caption{The two-point correlation function as a function of neutral fraction $\xh1$ for a given correlation length $r$ for the numerical simulation. Solid and dashed lines correspond to calculations with and without the redshift space distortion effect respectively. The magenta and blue lines show results for $\mu_r=1$ and $\mu_r=0$ respectively.  Here we have used the method described in Section \ref{Sec:coor-func} to perform the two-point correlation analysis on effectively 504 LC volumes thus suppressing the sample variance seen in Fig. \ref{fig:corr-func0}.}
\label{fig:corr_func1}
\end{figure*}

\begin{figure}
\includegraphics[width=0.5\textwidth]{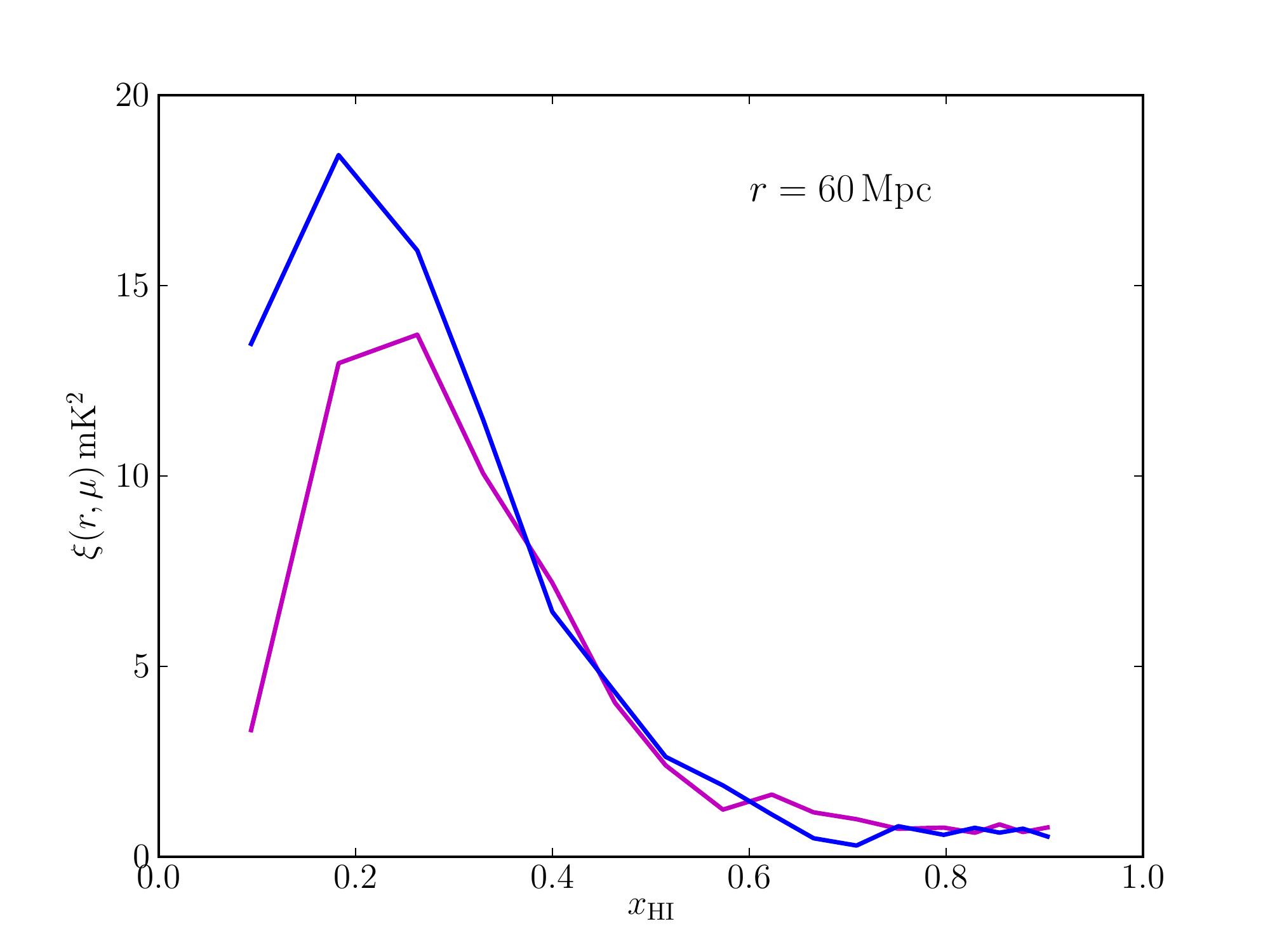}
\caption{The two-point correlation function  as a function of neutral fraction $\xh1$ for the semi-numerical simulation with out the redshift space distortion effect.   The magenta and blue lines show results for $\mu_r=1$ and $\mu_r=0$ respectively.}
\label{fig:corr_func2}
\end{figure}
Up to this point we have only considered the 21~cm power spectra in our analysis of the anisotropy due to the LC effect. However, a related but alternative quantity sometimes used for this is the two-point correlation function of the \hi 21-cm brightness temperature fluctuations. Traditionally, the correlation function has been used to study the large scale clustering of galaxies at lower redshifts \citep{1980lssu.book.....P, 1983ApJ...267..465D, 1993ApJ...412...64L}. There has not been much attention on using the correlation function to quantify the EoR \hi 21-cm signal, partly because the signal does not come from discrete sources but from the intergalactic medium. Another reason is that signal is observed in the Fourier domain and thus easy to analyze in the Fourier space. 

Nevertheless, \citet{2004ApJ...613....1F} developed analytic models to calculate the two-point correlation function for a given ionized bubble distribution. \citet{2005MNRAS.363..251A} used it to study the impact of the Alock-Paczynski effect and the redshift space distortion effect on EoR \hi 21-cm signal. \citet{barkana2006} used it to first study the LC anisotropy in the EoR \hi 21-cm signal. Recently \citet{2014arXiv1401.1807Z} used numerical simulations and studied the same aspects of the EoR and pre-EoR \hi 21-cm signal in more details. The two-point correlation function of the \hi 21-cm brightness temperature is defined as 
\be
\xi(r,\mu_r,z)=\langle(\delta T_{b,1}-\delta \bar{T}_b(z_1)) \times (\delta T_{b,2}-\delta \bar{T}_b(z_2)) \rangle.
\e 
Note that $\xi(r,\mu_r,z)$ is a function of the distance between two points $r$ and redshifts $z$. $\delta T_{b,i}$ is the \hi 21-cm brightness temperature at position $i$ corresponding to the redshift $z_i$. For the observed EoR \hi signal, the correlation function, in principle, will also change with the angle $\theta_r$ between the line-of-sight and the vector connecting the two points, where $\mu_r=\cos \theta_r$. $z$ is the redshift at the mid-point connecting the two points.

Calculating the correlation function for the entire 3D simulation box for all possible correlation lengths and $\mu_r$ is computationally expensive. The total number of operations to calculate the correlation function is $\sim N^2/2$ where $N$ is the total number of grid cells in the simulation. Instead, we calculate the correlation function for fixed correlation lengths. In addition, we restrict the correlation to a plane parallel ($\mu_r=1$) and perpendicular ($\mu_r=0$) to the line of sight. Fig. \ref{fig:corr-func0} shows the two-point correlation function as a function of neutral fraction $\xh1$ for a given correlation length $r$ for the numerical simulation. Here we only use our simulated light cone volume to calculate the correlation functions. For  $\mu_r=1$ we correlate points taken from two different 2D slices perpendicular to the line of sight which are respectively $r$ and $r/2$ distance apart from each other and from the centre corresponding to neutral fraction $x_{\rm HI}$. For $\mu_r=0$ we correlate points taken from the central slice at  neutral fraction $x_{\rm HI}$. We see that due to the small number of pairs available, the correlation function is strongly affected by the sample variance. This restricts us from deriving robust conclusions about the LC effect on the correlation function.

To suppress the sample variance we can construct multiple light cones from
the simulation. Since we can pick any two-dimensional slice to correspond to a
certain redshift, there are 504 different light cones that can be constructed
from the simulations data. However, since we do not need the full light cone
for the correlation function analysis, we employ a short cut using the coeval
volumes directly. To calculate the correlation function for the $\mu_r=1$ (line-of-sight) case and a specific correlation length $r$ we find pairs of coeval volumes $(z_1,z_2)$ whose difference in redshift $(z_2-z_1)$ corresponds to $r$. We also find $N_r$, the number of cells that corresponds to a distance $r$. Next we cross-correlate slice $n$ from coeval volume $z_1$ with slice $N_r+n$ from coeval volume $z_2$ for $n$ running from 1 to 504, $N=504$ is the number of cells in one spatial direction in the coeval volumes. In this procedure we use the periodic boundary conditions to handle the cases where $N_r+n > N$. For those cases we use slice number $N_r+n-N$. With this procedure we effectively use our complete data set to calculate the cross-correlations for $\mu_r=1$, namely 504 slices of each $(504)^2$ cells. This procedure is fully equivalent to constructing 504 light cones volumes without redshift space distortions and performing the correlation function analysis on these.

For the corresponding $\mu_r=0$ cross-correlations we select a coeval volume corresponding to a redshift $(z_1+z_2)/2$ and then calculate the correlation function of that coeval volume. 

For example, we have coeval 21-cm volumes at redshifts $z=7.059$ and $7.348$ which are a comoving distance $r \sim 100 \,{\rm Mpc}$ and $N_r\approx 83$ grid cells apart. So we cross-correlate slice $1$ from the simulation output at redshift $z=7.059$ with the $84th$ slice from the simulation output at redshift $z=7.348$, and slice $2$ with the $85$th slice and so on. 
For this case we use the coeval 21-com volume at $z=7.221$ to calculate the $\mu_r=0$ cross-correlations. With this procedure we effectively use our complete data set to calculate the cross-correlations for both values of $\mu_r$, namely 504 slices of each $(504)^2$ cells. A larger number of measurements will suppress the sample variance  considerably and we can study the correlation function at large length scales, where the signal is weak.

Fig. \ref{fig:corr_func1} shows the correlation function as a function of neutral fraction $\xh1$ for a given correlation length for the numerical simulation.  Here we see that the sample variance has been reduced considerably and we can study the LC effect on the correlation function. Dashed and solid lines correspond to without and with the redshift space distortion effect respectively. The magenta and blue lines show results for $\mu_r=1$ and $\mu_r=0$ respectively. We find that for the higher neutral fraction ($x_{\rm HI} \gtrsim 0.6$) the correlation function for $\mu_r=1$ (where the LC effect has been included) is the same as the corresponding $\mu_r=0$ lines even for a length scale of $r=100 \, {\rm Mpc}$. 

As reionization progresses, $\xi(r, \mu_r, z)$ for $\mu_r=1$ becomes lower than $\mu_r=0$. They again match with each other at the end of reionization. This trend is found to be the same even if we include the redshift space distortion effect (solid lines). Both cases peak essentially at the same phase of reionization at $x_{\rm HI}\approx 0.28$ for $r=100 \, {\rm Mpc}$ and $x_{\rm HI}\approx 0.25$ for $r=40 \, {\rm Mpc}$.

We find similar results for the semi-numerical simulation which represents the case of an early and rapid reionization (Fig. \ref{fig:corr_func2}). Unlike the numerical simulation, here the redshift space distortion effect hardly changes the results except at the beginning of the EoR and therefore we do not show them explicitly. In this case the $\mu_r=1$ case appears to peak slightly earlier. When comparing these results to those in \cite{barkana2006} and \citet{2014arXiv1401.1807Z} several similarities and differences can be noted. As in \cite{barkana2006} we find that the peak of $\xi(\mu_r=0)$ is the higher one. We also agree with those authors on the stage of reionization when the largest differences occur. However, in \cite{barkana2006} the peaks are much more clearly separated (see their fig.~4), whereas we only find a small difference for the semi-numerical simulation and none for the numerical one. \citet{2014arXiv1401.1807Z} show that $\xi(\mu_r=1)$ peaks twice, once before and once after the peak in $\xi(\mu_r=0)$ (see their fig.~6). In contrast with both \cite{barkana2006} and our results, the peaks are highest in the  $\xi(\mu_r=1)$ case. All results agree that one needs to look at large scales ($\sim 100 \, {\rm Mpc}$) to see differences between the parallel and perpendicular correlation functions.

It is not obvious what causes these differences. \citet{barkana2006} used very simplified analytic models. In the simulations of \citet{2014arXiv1401.1807Z}, low-mass haloes are missing. Very massive and luminous sources contribute to the reionization and hence the reionization history is very different from our simulations. Their results appear sample variance dominated, as the lines are not very smooth. This might explain the differences between our work and previous works. Another notable difference is that for our numerical result the signal strength is much smaller. For example, the peak values for $\xi(r, \mu_r, z)$  are $\sim 0.4 \, {\rm mK^2}$ and $\sim 4 \, {\rm mK^2}$ for the correlation length $r=100$ and $40 \, {\rm Mpc}$. The above two papers find $5-10$ times stronger signals. Small ionized regions and the resulting low variance of the 21 cm signal in our simulations compared to the other two could explain this.

The  power spectrum  and two-point correlation function  contain the same basic information about the structure in the observed 21-cm brightness temperature field. So it must be true that they are either both isotropic or both anisotropic. However, the power spectra calculated for a fixed $k$-mode for different $\mu_r$-values (see Figures \ref{fig:aniso-c2ray1}, \ref{fig:aniso-semi-num})  are similar. On the other hand the correlation function calculated for a fixed length scale $r$ for different $\mu_r$-values are found to be different. 

The apparent inconsistency probably arises due to the way the correlation function is calculated. Like \cite{barkana2006} and \citet{2014arXiv1401.1807Z}, we calculate the correlation function for $\mu_r=1$ (the plane that is parallel to the line of sight) by only considering those pairs whose middle point corresponds to a specific central redshift. For example, let us consider a light cone volume which runs from $z_\mathrm{start}$ to $z_\mathrm{end}$, corresponding to a comoving distance $L_\mathrm{LC}$. If we want to measure the correlation function of this volume for a distance $L$ ($< L_\mathrm{LC}$) we should find all pairs of slices along the line-of-sight which are a distance $L$ apart. The value of this correlation function is the Fourier transform of the power spectrum measured at a scale $k=2\pi/L$. However, this value is calculated from many different pairs of redshifts and mixes the signal from different phases of reionization. If we instead choose a specific pair of redshifts, $z_1$ and $z_2$ which are a distance $L$ apart (with a central redshift $z_\mathrm{c}$), we obtain a correlation function which does no longer correspond to the Fourier transform of the power spectrum measured at a scale $k=2\pi/L$ but which does measure the correlation at a specific phase of reionization. It is the latter quantity which we have studied in this section and appears to be more useful for picking out the anisotropy in light cone data.

Finally we note that the correlation function considered here is a measurable quantity. For a given redshift $z_c$ and distance $r$ one picks images a distance $r$ apart and straddling this redshift $z_c$. This way the correlation measurement belongs to a clear redshift, important for characterizing the light cone effect. This method can be applied in a straightforward way to real data. However, as can be seen in Fig. \ref{fig:corr-func0} or in \citet{2014arXiv1401.1807Z} for one field of view of several degrees across this measurement will be swamped by sample variance. Instruments like LOFAR and SKA have fields-of-view of about $\sim 5\degr \times 5\degr$, comparable to the simulation size we consider here. The simulation results indicate that several hundreds of these would be needed to determine the cross-correlation with minimal sample variance. Given that these fields each need $\sim 1000$ of hours of observation time, such numbers of fields will not be available. Thus, for LOFAR and SKA, this measurement is not actually feasible. However, experiments with larger field-of-view, such as MWA or PAPER, would only need tens of different patches which would allow a cross-correlation analysis, although it may be practically challenging.

\section{Optimum bandwidth for analyzing observed data}
\label{sec:optimum-bw}
\begin{figure*}
\includegraphics[width=0.8\textwidth]{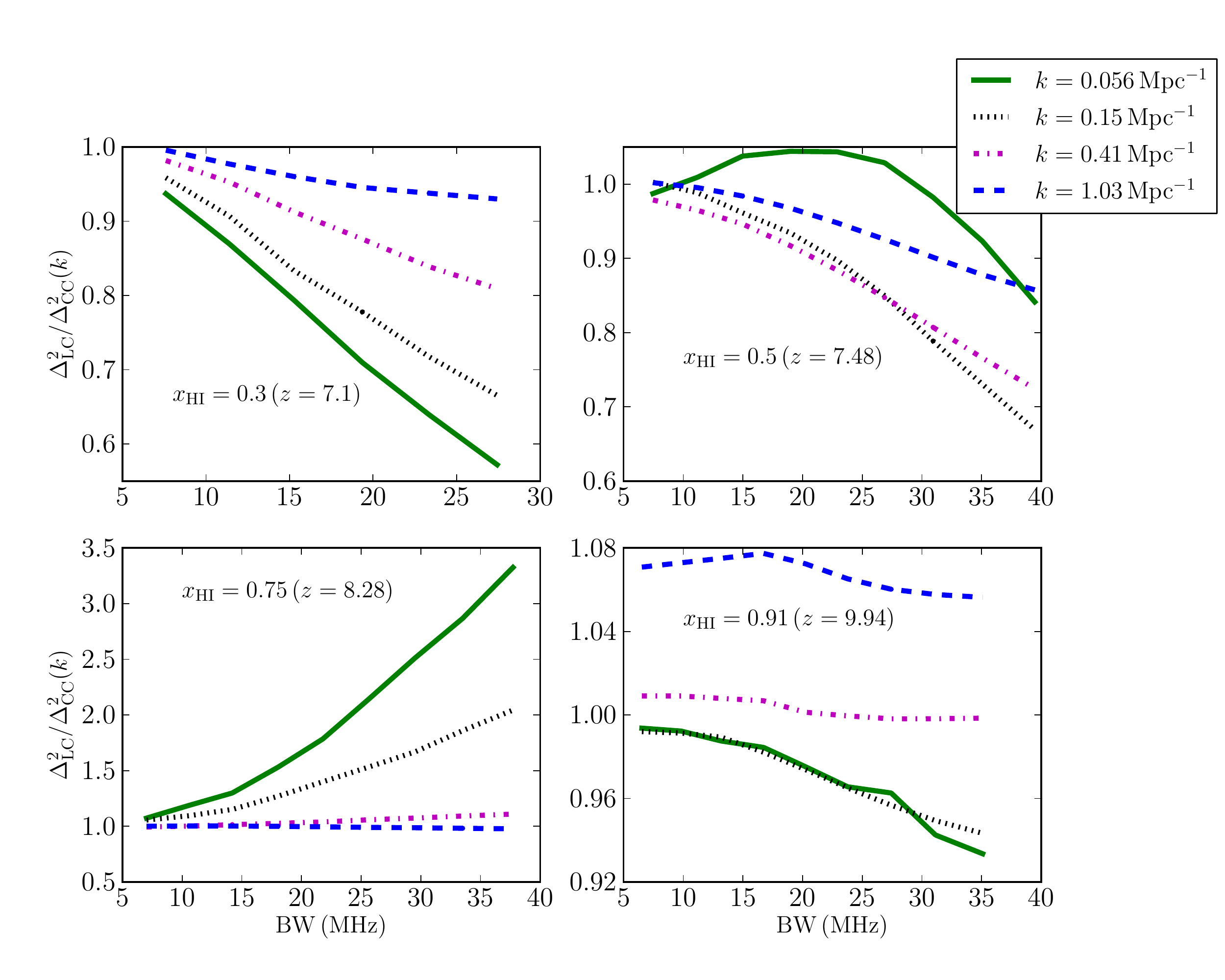}
\caption{The ratio between power spectra with and without the LC effect as a function frequency bandwidth over which the power spectra have been evaluated. Note that the scale on the $y$-axis is different for the different panels.}
\label{fig:ratio-bw}
\end{figure*}

\begin{figure*}
\includegraphics[width=0.8\textwidth]{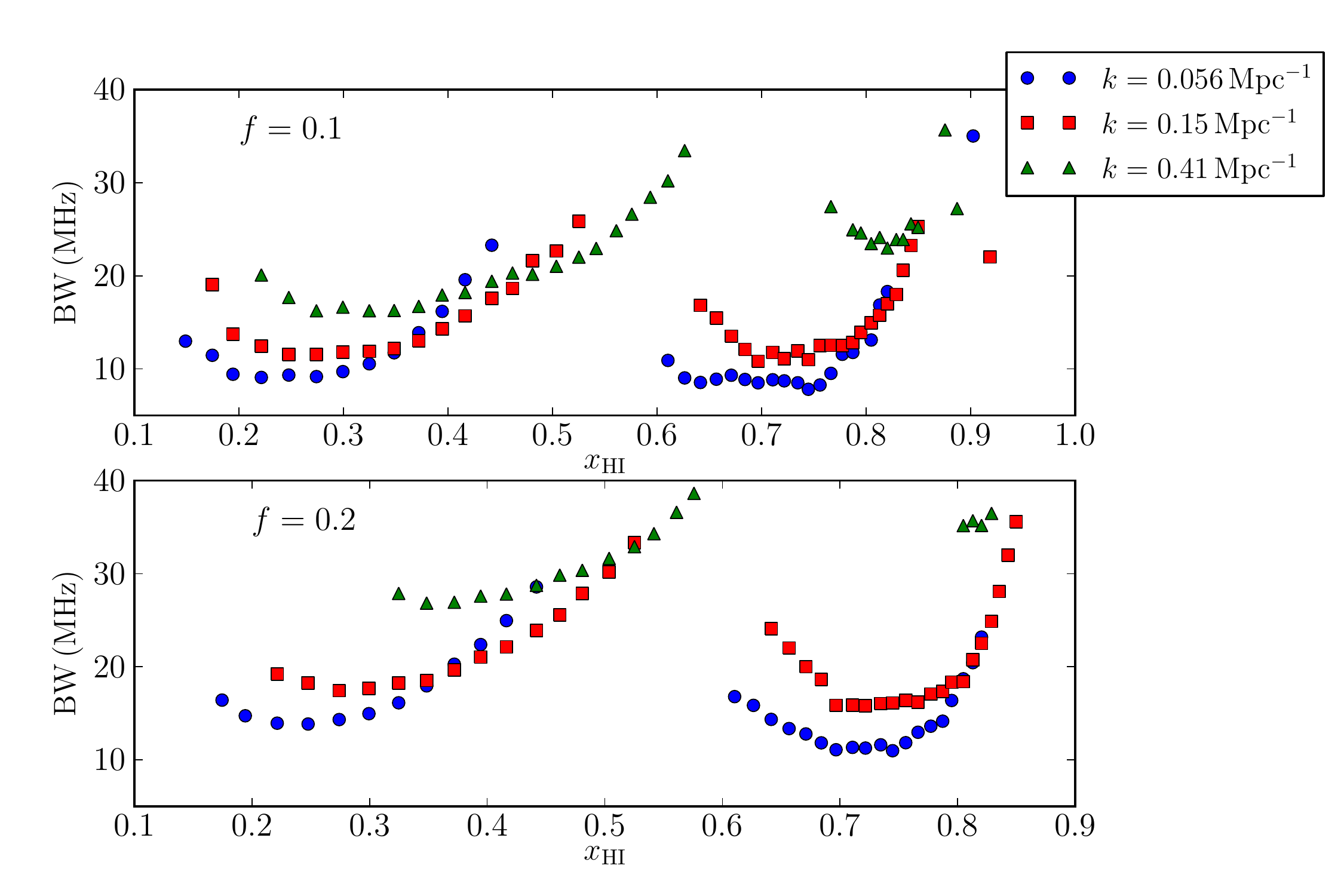}
\caption{The optimum bandwidth for analyzing observed EoR \hi 21-cm data as a function of the neutral fraction $\xh1$ for different $k$-modes. The gap in each line around $\xh1 \sim 0.5$ to $0.7$ indicates that the optimum bandwidth for those neutral fractions are much higher that full box size.}
\label{fig:bw-redshift}
\end{figure*}
When analyzing the 21~cm power spectra above in Sect.~\ref{sect:sph_avg_ps} we  consider two different values for the length of the line-of-sight (see Figures \ref{fig:ratio} and \ref{fig:ratio-pv}). We find that the LC effect on the \hi 21-cm power spectrum is smaller for the shorter line-of-sight. It would be interesting to investigate how the LC effect changes as we change the length of our line-of-sight. Obviously as we reduce the line-of-sight, the power spectra with and without the LC effect will start to converge.

In Fig. \ref{fig:ratio-bw} we plot the ratio between the spherically-averaged power spectra with and without the LC effect, $\Delta_{\rm{LC}}^2(k)/\Delta_{\rm{CC}}^2(k)$, as a function of the frequency bandwidth or equivalently the line-of-sight length for different $k$-modes and at different redshifts.  The results are shown in redshift space, i.e. we incorporate the redshift space distortion effect in both the power spectra with and without the LC effect as it will be observed by an instrument. 

In general, we find that the ratio deviates from $1$ more for the higher bandwidth. For the neutral fraction $x_{\rm HI}=0.3$ $(z=7.1)$ (top left panel) the ratio decreases with increasing value of the bandwidth. For example, for $k=0.056 \, \mpci$  the ratio is about $0.55$ at bandwidth$=27$ MHz. The ratio gradually approaches $1$ as we decrease the bandwidth. At $10$ MHz  the ratio comes close to $0.9$. We see similar trends for other $k$ modes.  For $\xh1=0.5$ (top right panel) the ratio for $k=0.056$ is very close to $1$ even at larger bandwidth, consistent with our previous results. For $k=0.15$ and $0.41  \, \mpci$ the ratio departs considerably from $1$ at the larger bandwidth. Around $\xh1=0.75 $ (bottom left panel) the ratio is always greater than $1$ and very high for $k=0.056$ and $0.15\, \mpci$ consistent with our results (see Fig. \ref{fig:ratio-pv}). The ratio gradually decreases with bandwidth and converges to the value $1$ as we decrease the bandwidth. In the early phases of the EoR the LC effect is small and hence the ratio is always very close to $1$ and the deviation is less than $10\%$ for all $k$-modes (see the bottom right panel).

It is now obvious that the LC effect should be considered if we estimate the EoR  \hi 21-cm power spectrum for longer line-of-sights, that is for very large frequency bandwidths, otherwise the predictions will be wrong or the data would be mis-interpreted. Taking care of the LC effect properly may not be straightforward as the effect is highly model dependent. To avoid any complications that may arise due the uncertainty of modeling the LC effect, data analysis over smaller bandwidth is typically preferred (e.g, \citet{2010MNRAS.405.2492H}). On the other hand, a large bandwidth is required to reduce the instrumental system noise and increase the sensitivity for detecting the signal. A larger bandwidth will also allow us to probe large line-of-sight scales, i.e. small $\kpl$ \citep{mcquinn2006}. 

Therefore, the optimum strategy would be to use a bandwidth as large as possible so that large scale line-of-sight distances can be probed but still keeping the LC effect negligible. Therefor, prior knowledge about the optimum bandwidth would be useful for analyzing the observed EoR \hi 21-cm data where the effect can be neglected. We now try to find this optimum bandwidth. In Fig. \ref{fig:bw-redshift} we plot the bandwidth which allows a fractional change $f$ in the power spectrum with the LC effect with respect to the power spectrum without the LC effect as a function of the neutral fraction $\xh1$. We define $f$ as:
\be
f=\left | \frac{\Delta_{\rm LC}^2(k)}{\Delta_{\rm CC}^2(k)}-1 \right|.
\e
The upper panel shows results for $f=0.1$ for three different $k$-modes. This means that we calculate the bandwidth (or equivalently the line-of-sight width) for which the power spectrum with the LC effect deviates $\leq 10\%$ from the power spectrum without the effect. The gaps in the lines around $\xh1 \sim 0.5$ to $0.7$ indicate that the optimum bandwidth for this neutral fraction range are much higher than the extent of our full box size. We find that the optimum bandwidth changes with $\xh1$ and the minimum optimum bandwidth for $k=0.056 \, \rm{Mpc}^{-1}$ is $\sim 7.5 $ MHz. For $k=0.15$ and $0.41 \, \rm{Mpc}^{-1}$ the optimum bandwidth is $\sim 11$ and $\sim 16$ MHz respectively. 

If we allow the fractional change to be $f=0.2$ (see lower panel) the optimum bandwidth becomes higher: $\sim 11$, $\sim 16$ and $27$ MHz for $k=0.056$, $0.15$ and $0.41 \, \rm{Mpc}^{-1}$ respectively. Our results suggest that if one wants to analyze the power spectrum at modes $k \gtrsim 0.056 \, \mpci$ and one is ready to neglect $10\%$ error in the power spectrum, the data should be analyzed over the bandwidth $\lesssim 7.5$ MHz. For higher $k$-modes and larger fractional change $f$, a larger optimum bandwidth is allowed. The above estimate will change for different reionization models. Since in our numerical simulation the reionization is more gradual and extended, the LC effect is smaller, and therefore our estimate gives a rough idea about the optimum bandwidth and should be considered as an upper limit.

\section{Summary and Discussion}
We use the largest size ($607 \, \mpc$) radiative transfer simulation of the EoR to date to investigate the impact of the light cone (LC) effect on the EoR \hi 21-cm signal. In particular, we focus on the evolution of the effect at different scales during the EoR, anisotropies in the \hi 21-cm signal and observational implications. We present results both in real and redshift space by including the effects of peculiar velocities. 

We find that the LC effect is most dramatic at two different stages of the EoR: one when reionization is $\sim 20\%$ and other when it is $\sim 80\%$ finished. The effect is relatively small at $\sim 50\%$ reionization, consistent with our previous results. We argue that the non-linear evolution of the power spectrum as a function of comoving distance from the present determines the light cone effect where any linear evolution gets averaged out. We observe up to $\sim 40\%$ and $\sim 25\%$ change at the largest scale available for a line-of-sight length of $241 \, \mpc$. When we use the entire simulation box ($607 \, \mpc$) we find a factor of $\sim 4$ amplification in the power spectrum at $\xh1 \sim 0.75$. 

We find some similarities between the ways the LC and the redshift space distortion effect affect the spherically-averaged \hi 21-cm power spectrum. For example, both effects enhance the power spectra significantly at large scales and at the initial stages ($\xh1 \sim 0.8$) of the EoR. However the redshift space distortion effect becomes very small at the late stages of the EoR ($\xh1 \lesssim 0.3$) unlike the LC effect which also has a significant effect at the late stages. Unlike the redshift space distortion effect, the LC effect depends on the length of the line-of-sight used in the analysis. 

Somewhat surprisingly, we do not observe any significant anisotropy in the \hi 21-cm power spectrum due to the LC effect. We find similar results in our semi-numerical simulation of an early rapid reionization. The models used in this paper did not result in relativistically propagating ionization fronts and therefore the ionized bubbles do not get elongated or compressed systematically due to the LC effect. They only appear smaller or bigger without changing the shape drastically. Using relevant toy models we argue that the systematic change in the bubble sizes as a function of the line-of-sight position while individual bubbles remain spherical is not enough to make the power spectrum anisotropic. Individual bubbles should be either elongated or compressed to make the power spectrum anisotropic. The reason that ionized bubbles in our simulations do not get elongated or compressed is that sources are not bright enough to make the ionization font relativistic which is an important property to make bubbles elongated or compressed.

We also calculate the two-point correlation function of the \hi 21-cm brightness temperature fluctuations. The two-point correlation function for a given length scale peaks around $\xh1 \sim 0.3$. When we calculate the correlation functions along and perpendicular to the line-of-sight they are found to be different for scales of $\sim 100 \mpc$ and late stages of reionization. This is consistent with earlier works. This apparent `anisotropy' found in the correlation function is due to differences in the way we calculate the two correlation functions along and perpendicular to the line-of-sight. We consider all possible correlations when we calculate it for perpendicular to the line-of-sight whereas we impose some restrictions while calculating it along the line-of-sight. However, for small scales ($\lesssim 40 \, \mpc$) and in the early stages the parallel and perpenducular correlation functions are identical.

The LC effect is obviously dependent on the line-of-sight extent over which the signal is being analyzed. The smaller the line-of-sight extent or frequency bandwidth, the smaller the effect. We go on to calculate the `optimum bandwidth' for predicting or analyzing EoR \hi 21-cm power spectra. The optimum bandwidth is the highest bandwidth allowed for analyzing the EoR \hi 21-cm power spectra for which the LC effect can be neglected. We find that the optimum bandwidth for $k= 0.056 \, \mpci$ is $\sim 7.5$ MHz. For $k = 0.15$ and $0.41 \, \mpci$ the optimum bandwidth is $\sim 11$ and $\sim 16$ MHz respectively if we can neglect a $10\%$ change in the power spectra.

These results might change for different reionization models. Our reionization model is  extended (extended over $\Delta z \approx 3$) and sources are relatively weak. Therefore, we think that the LC effect presented here is on the lower side. Therefore the optimum bandwidth would be smaller for more rapid reionization models. However, we believe that our results provide a reasonable estimate of the effect. The signal is also expected to be isotropic or weakly anisotropic, unless reionization is dominated by very bright sources like super-massive quasars. The relatively high value of the electron scattering optical depth combined with the measurements of the UV background around $z\sim 6$ do suggest a more extended reionization.

It has been proposed that the EoR H~I 21-cm signal can be used to extract the `pure' dark matter power spectrum through a fitting of the anisotropic power spectrum \citep{barkana2006, 2013PhRvL.110o1301S}. In addition the signal contains information about the correlation between the distribution of the sources of radiation and the matter density and can thus be used to characterize the sources of reionization \citep{2013MNRAS.435..460J}. The impact of the light cone effect on the above issues has not been investigated in details although \citet{2013MNRAS.435..460J} included the effect. Our results show that the LC effect does not introduce any anisotropy in the power spectrum, which suggests that ignoring the LC effect in such studies is probably valid.

In this paper we only considered the high spin temperature case for which the H~I 21-cm signal appears in emission. However, during the pre-reionization phase spin temperature variations may cause the signal to appear in absorption and results may differ from what we find here \citep{2014arXiv1401.1807Z}. We would like to investigate this in more detail in future.

\section*{Acknowledgments} 
KKD thanks the Department of Science \& Technology (DST), India for
the research grant SR/FTP/PS-119/2012 under the Fast Track Scheme for
Young Scientist. KKD is grateful for financial support from Swedish
Research Council (VR) through the Oscar Klein Centre (grant
2007-8709). KKD would like to thank Tirthankar Roy Choudhury and
Raghunath Ghara for useful discussion. GM is supported by Swedish
Research Council grant 2012-4144. YM was supported by French state
funds managed by the ANR within the Investissements d'Avenir programme
under reference ANR-11-IDEX-0004-02. PRS was supported in part by
U.S. NSF grants AST-0708176 and AST-1009799, and NASA grants
NNX07AH09G and NNX11AE09G. KA was supported in part by NRF grant funded by the Korean government MEST (No. 2012R1A1A1014646).
This work was supported by the Science and 
Technology Facilities Council [grant numbers ST/F002858/1 and ST/I000976/1]; 
and The Southeast Physics Network (SEPNet). 
The authors acknowledge the U.S. NSF
TeraGrid/XSEDE Project AST090005 and the Texas Advanced
Computing Center (TACC) at The University of Texas at Austin for providing
HPC resources that have contributed to the research results reported within
this paper. This research was supported in part by an allocation of advanced
computing resources provided by the National Science Foundation through TACC
and the National Institute for Computational Sciences (NICS), with part of the
computations performed on Lonestar at TACC (http://www.tacc.utexas.edu) and
Kraken at NICS (http://www.nics.tennessee.edu/). Some of the numerical
computations were done on the Apollo cluster at The University of Sussex and
the Sciama High Performance Compute (HPC) cluster which is supported by the ICG,
SEPNet and the University of Portsmouth. Part of the computations were performed
on the GPC supercomputer at the SciNet HPC Consortium. SciNet is funded by: the
Canada Foundation for Innovation under the auspices of Compute Canada; the
Government of Ontario; Ontario Research Fund - Research Excellence; and the
University of Toronto.

\bibliography{lc-draft-version2.bib}

\begin{thebibliography}{75}
\expandafter\ifx\csname natexlab\endcsname\relax\def\natexlab#1{#1}\fi

\bibitem[{{Alcock} \& {Paczynski}(1979)}]{1979Natur.281..358A}
{Alcock} C., {Paczynski} B., 1979, \nat, 281, 358

\bibitem[{{Ali}, {Bharadwaj} \& {Chengalur}(2008){Ali}, {Bharadwaj}, \&
  {Chengalur}}]{2008MNRAS.385.2166A}
{Ali} S.~S., {Bharadwaj} S., {Chengalur} J.~N., 2008, \mnras, 385, 2166

\bibitem[{{Ali}, {Bharadwaj} \& {Pandey}(2005){Ali}, {Bharadwaj}, \&
  {Pandey}}]{2005MNRAS.363..251A}
{Ali} S.~S., {Bharadwaj} S., {Pandey} B., 2005, \mnras, 363, 251

\bibitem[{{Baek} {et~al}\mbox{.}(2009){Baek}, {Di Matteo}, {Semelin}, {Combes},
  \& {Revaz}}]{2009A&A...495..389B}
{Baek} S., {Di Matteo} P., {Semelin} B., {Combes} F., {Revaz} Y., 2009, \aap,
  495, 389

\bibitem[{{Barkana}(2006)}]{2006MNRAS.372..259B}
{Barkana} R., 2006, \mnras, 372, 259

\bibitem[{{Barkana} \& {Loeb}(2005)}]{2005ApJ...624L..65B}
{Barkana} R., {Loeb} A., 2005, \apjl, 624, L65

\bibitem[{{Barkana} \& {Loeb}(2006)}]{barkana2006}
{Barkana} R., {Loeb} A., 2006, \mnras, 372, L43

\bibitem[{{Battaglia} {et~al}\mbox{.}(2013){Battaglia}, {Trac}, {Cen}, \&
  {Loeb}}]{2013ApJ...776...81B}
{Battaglia} N., {Trac} H., {Cen} R., {Loeb} A., 2013, \apj, 776, 81

\bibitem[{{Bernardi} {et~al}\mbox{.}(2009){Bernardi}, {de Bruyn}, {Brentjens},
  {Ciardi}, {Harker}, {Jeli{\'c}}, {Koopmans}, {Labropoulos}, {Offringa},
  {Pandey}, {Schaye}, {Thomas}, {Yatawatta}, \&
  {Zaroubi}}]{2009A&A...500..965B}
{Bernardi} G. {et~al.}, 2009, \aap, 500, 965

\bibitem[{{Bernardi} {et~al}\mbox{.}(2013){Bernardi}, {Greenhill}, {Mitchell},
  {Ord}, {Hazelton}, {Gaensler}, {de Oliveira-Costa}, {Morales}, {Udaya
  Shankar}, {Subrahmanyan}, {Wayth}, {Lenc}, {Williams}, {Arcus}, {Arora},
  {Barnes}, {Bowman}, {Briggs}, {Bunton}, {Cappallo}, {Corey}, {Deshpande},
  {deSouza}, {Emrich}, {Goeke}, {Herne}, {Hewitt}, {Johnston-Hollitt},
  {Kaplan}, {Kasper}, {Kincaid}, {Koenig}, {Kratzenberg}, {Lonsdale}, {Lynch},
  {McWhirter}, {Morgan}, {Oberoi}, {Pathikulangara}, {Prabu}, {Remillard},
  {Rogers}, {Roshi}, {Salah}, {Sault}, {Srivani}, {Stevens}, {Tingay},
  {Waterson}, {Webster}, {Whitney}, {Williams}, \&
  {Wyithe}}]{2013ApJ...771..105B}
{Bernardi} G. {et~al.}, 2013, \apj, 771, 105

\bibitem[{{Bharadwaj} \& {Ali}(2005)}]{2005MNRAS.356.1519B}
{Bharadwaj} S., {Ali} S.~S., 2005, \mnras, 356, 1519

\bibitem[{{Bittner} \& {Loeb}(2011)}]{2011JCAP...04..038B}
{Bittner} J.~M., {Loeb} A., 2011, \jcap, 4, 38

\bibitem[{{Bowman} \& {Rogers}(2010)}]{2010Natur.468..796B}
{Bowman} J.~D., {Rogers} A.~E.~E., 2010, \nat, 468, 796

\bibitem[{{Chapman} {et~al}\mbox{.}(2013){Chapman}, {Abdalla}, {Bobin},
  {Starck}, {Harker}, {Jeli{\'c}}, {Labropoulos}, {Zaroubi}, {Brentjens}, {de
  Bruyn}, \& {Koopmans}}]{2013MNRAS.429..165C}
{Chapman} E. {et~al.}, 2013, \mnras, 429, 165

\bibitem[{{Choudhury}, {Haehnelt} \& {Regan}(2009){Choudhury}, {Haehnelt}, \&
  {Regan}}]{2009MNRAS.394..960C}
{Choudhury} T.~R., {Haehnelt} M.~G., {Regan} J., 2009, \mnras, 394, 960

\bibitem[{{Datta}, {Bharadwaj} \& {Choudhury}(2007){Datta}, {Bharadwaj}, \&
  {Choudhury}}]{2007MNRAS.382..809D}
{Datta} K.~K., {Bharadwaj} S., {Choudhury} T.~R., 2007, \mnras, 382, 809

\bibitem[{{Datta} {et~al}\mbox{.}(2012{\natexlab{a}}){Datta}, {Friedrich},
  {Mellema}, {Iliev}, \& {Shapiro}}]{2012MNRAS.424..762D}
{Datta} K.~K., {Friedrich} M.~M., {Mellema} G., {Iliev} I.~T., {Shapiro} P.~R.,
  2012{\natexlab{a}}, \mnras, 424, 762

\bibitem[{{Datta} {et~al}\mbox{.}(2012{\natexlab{b}}){Datta}, {Mellema}, {Mao},
  {Iliev}, {Shapiro}, \& {Ahn}}]{2012MNRAS.424.1877D}
{Datta} K.~K., {Mellema} G., {Mao} Y., {Iliev} I.~T., {Shapiro} P.~R., {Ahn}
  K., 2012{\natexlab{b}}, \mnras, 424, 1877

\bibitem[{{Davis} \& {Peebles}(1983)}]{1983ApJ...267..465D}
{Davis} M., {Peebles} P.~J.~E., 1983, \apj, 267, 465

\bibitem[{{Furlanetto}, {Zaldarriaga} \& {Hernquist}(2004){Furlanetto},
  {Zaldarriaga}, \& {Hernquist}}]{2004ApJ...613....1F}
{Furlanetto} S.~R., {Zaldarriaga} M., {Hernquist} L., 2004, \apj, 613, 1

\bibitem[{{Geil} \& {Wyithe}(2008)}]{2008MNRAS.386.1683G}
{Geil} P.~M., {Wyithe} J.~S.~B., 2008, \mnras, 386, 1683

\bibitem[{{Ghosh} {et~al}\mbox{.}(2012){Ghosh}, {Prasad}, {Bharadwaj}, {Ali},
  \& {Chengalur}}]{2012MNRAS.426.3295G}
{Ghosh} A., {Prasad} J., {Bharadwaj} S., {Ali} S.~S., {Chengalur} J.~N., 2012,
  \mnras, 426, 3295

\bibitem[{{Harker} {et~al}\mbox{.}(2010){Harker}, {Zaroubi}, {Bernardi},
  {Brentjens}, {de Bruyn}, {Ciardi}, {Jeli{\'c}}, {Koopmans}, {Labropoulos},
  {Mellema}, {Offringa}, {Pandey}, {Pawlik}, {Schaye}, {Thomas}, \&
  {Yatawatta}}]{2010MNRAS.405.2492H}
{Harker} G. {et~al.}, 2010, \mnras, 405, 2492

\bibitem[{{Harker} {et~al}\mbox{.}(2009){Harker}, {Zaroubi}, {Thomas},
  {Jeli{\'c}}, {Labropoulos}, {Mellema}, {Iliev}, {Bernardi}, {Brentjens}, {de
  Bruyn}, {Ciardi}, {Koopmans}, {Pandey}, {Pawlik}, {Schaye}, \&
  {Yatawatta}}]{2009MNRAS.393.1449H}
{Harker} G.~J.~A. {et~al.}, 2009, \mnras, 393, 1449

\bibitem[{{Harnois-D{\'e}raps} {et~al}\mbox{.}(2013){Harnois-D{\'e}raps},
  {Pen}, {Iliev}, {Merz}, {Emberson}, \& {Desjacques}}]{2013MNRAS.436..540H}
{Harnois-D{\'e}raps} J., {Pen} U.-L., {Iliev} I.~T., {Merz} H., {Emberson}
  J.~D., {Desjacques} V., 2013, \mnras, 436, 540

\bibitem[{{Hinshaw} {et~al}\mbox{.}(2013){Hinshaw}, {Larson}, {Komatsu},
  {Spergel}, {Bennett}, {Dunkley}, {Nolta}, {Halpern}, {Hill}, {Odegard},
  {Page}, {Smith}, {Weiland}, {Gold}, {Jarosik}, {Kogut}, {Limon}, {Meyer},
  {Tucker}, {Wollack}, \& {Wright}}]{2013ApJS..208...19H}
{Hinshaw} G. {et~al.}, 2013, \apjs, 208, 19

\bibitem[{{Iliev} {et~al}\mbox{.}(2006){Iliev}, {Ciardi}, {Alvarez}, {Maselli},
  {Ferrara}, {Gnedin}, {Mellema}, {Nakamoto}, {Norman}, {Razoumov},
  {Rijkhorst}, {Ritzerveld}, {Shapiro}, {Susa}, {Umemura}, \&
  {Whalen}}]{2006MNRAS.371.1057I}
{Iliev} I.~T. {et~al.}, 2006, \mnras, 371, 1057

\bibitem[{{Iliev} {et~al}\mbox{.}(2014){Iliev}, {Mellema}, {Ahn}, {Shapiro},
  {Mao}, \& {Pen}}]{2014MNRAS.439..725I}
{Iliev} I.~T., {Mellema} G., {Ahn} K., {Shapiro} P.~R., {Mao} Y., {Pen} U.-L.,
  2014, \mnras, 439, 725

\bibitem[{{Iliev} {et~al}\mbox{.}(2007){Iliev}, {Mellema}, {Shapiro}, \&
  {Pen}}]{2007MNRAS.376..534I}
{Iliev} I.~T., {Mellema} G., {Shapiro} P.~R., {Pen} U.-L., 2007, \mnras, 376,
  534

\bibitem[{{Jacobs} {et~al}\mbox{.}(2013){Jacobs}, {Parsons}, {Aguirre}, {Ali},
  {Bowman}, {Bradley}, {Carilli}, {DeBoer}, {Dexter}, {Gugliucci}, {Klima},
  {MacMahon}, {Manley}, {Moore}, {Pober}, {Stefan}, \&
  {Walbrugh}}]{2013ApJ...776..108J}
{Jacobs} D.~C. {et~al.}, 2013, \apj, 776, 108

\bibitem[{{Jeli{\'c}} {et~al}\mbox{.}(2008){Jeli{\'c}}, {Zaroubi},
  {Labropoulos}, {Thomas}, {Bernardi}, {Brentjens}, {de Bruyn}, {Ciardi},
  {Harker}, {Koopmans}, {Pandey}, {Schaye}, \&
  {Yatawatta}}]{2008MNRAS.389.1319J}
{Jeli{\'c}} V. {et~al.}, 2008, \mnras, 389, 1319

\bibitem[{{Jensen} {et~al}\mbox{.}(2013){Jensen}, {Datta}, {Mellema},
  {Chapman}, {Abdalla}, {Iliev}, {Mao}, {Santos}, {Shapiro}, {Zaroubi},
  {Bernardi}, {Brentjens}, {de Bruyn}, {Ciardi}, {Harker}, {Jeli{\'c}},
  {Kazemi}, {Koopmans}, {Labropoulos}, {Martinez}, {Offringa}, {Pandey},
  {Schaye}, {Thomas}, {Veligatla}, {Vedantham}, \&
  {Yatawatta}}]{2013MNRAS.435..460J}
{Jensen} H. {et~al.}, 2013, \mnras, 435, 460

\bibitem[{{La Plante} {et~al}\mbox{.}(2013){La Plante}, {Battaglia},
  {Natarajan}, {Peterson}, {Trac}, {Cen}, \& {Loeb}}]{plante2013}
{La Plante} P., {Battaglia} N., {Natarajan} A., {Peterson} J.~B., {Trac} H.,
  {Cen} R., {Loeb} A., 2013, arxiv-1309.7056

\bibitem[{{Landy} \& {Szalay}(1993)}]{1993ApJ...412...64L}
{Landy} S.~D., {Szalay} A.~S., 1993, \apj, 412, 64

\bibitem[{{Majumdar}, {Bharadwaj} \& {Choudhury}(2012){Majumdar}, {Bharadwaj},
  \& {Choudhury}}]{2012MNRAS.426.3178M}
{Majumdar} S., {Bharadwaj} S., {Choudhury} T.~R., 2012, \mnras, 426, 3178

\bibitem[{{Majumdar}, {Bharadwaj} \& {Choudhury}(2013){Majumdar}, {Bharadwaj},
  \& {Choudhury}}]{2013MNRAS.434.1978M}
{Majumdar} S., {Bharadwaj} S., {Choudhury} T.~R., 2013, \mnras, 434, 1978

\bibitem[{{Majumdar} {et~al}\mbox{.}(2011){Majumdar}, {Bharadwaj}, {Datta}, \&
  {Choudhury}}]{2011MNRAS.413.1409M}
{Majumdar} S., {Bharadwaj} S., {Datta} K.~K., {Choudhury} T.~R., 2011, \mnras,
  413, 1409

\bibitem[{{Majumdar} {et~al}\mbox{.}(2014){Majumdar}, {Mellema}, {Datta},
  {Jensen}, {Choudhury}, {Bharadwaj}, \& {Friedrich}}]{2014arXiv1403.0941M}
{Majumdar} S., {Mellema} G., {Datta} K.~K., {Jensen} H., {Choudhury} T.~R.,
  {Bharadwaj} S., {Friedrich} M.~M., 2014, ArXiv e-prints, arXiv: 1403.0941

\bibitem[{{Malloy} \& {Lidz}(2013)}]{2013ApJ...767...68M}
{Malloy} M., {Lidz} A., 2013, \apj, 767, 68

\bibitem[{{Mao} {et~al}\mbox{.}(2012){Mao}, {Shapiro}, {Mellema}, {Iliev},
  {Koda}, \& {Ahn}}]{2012MNRAS.422..926M}
{Mao} Y., {Shapiro} P.~R., {Mellema} G., {Iliev} I.~T., {Koda} J., {Ahn} K.,
  2012, \mnras, 422, 926

\bibitem[{{Maselli} {et~al}\mbox{.}(2007){Maselli}, {Gallerani}, {Ferrara}, \&
  {Choudhury}}]{2007MNRAS.376L..34M}
{Maselli} A., {Gallerani} S., {Ferrara} A., {Choudhury} T.~R., 2007, \mnras,
  376, L34

\bibitem[{{McQuinn} {et~al}\mbox{.}(2007){McQuinn}, {Lidz}, {Zahn}, {Dutta},
  {Hernquist}, \& {Zaldarriaga}}]{2007MNRAS.377.1043M}
{McQuinn} M., {Lidz} A., {Zahn} O., {Dutta} S., {Hernquist} L., {Zaldarriaga}
  M., 2007, \mnras, 377, 1043

\bibitem[{{McQuinn} {et~al}\mbox{.}(2006){McQuinn}, {Zahn}, {Zaldarriaga},
  {Hernquist}, \& {Furlanetto}}]{mcquinn2006}
{McQuinn} M., {Zahn} O., {Zaldarriaga} M., {Hernquist} L., {Furlanetto} S.~R.,
  2006, \apj, 653, 815

\bibitem[{{Mellema} {et~al}\mbox{.}(2006{\natexlab{a}}){Mellema}, {Iliev},
  {Alvarez}, \& {Shapiro}}]{2006NewA...11..374M}
{Mellema} G., {Iliev} I.~T., {Alvarez} M.~A., {Shapiro} P.~R.,
  2006{\natexlab{a}}, \na, 11, 374

\bibitem[{{Mellema} {et~al}\mbox{.}(2006{\natexlab{b}}){Mellema}, {Iliev},
  {Pen}, \& {Shapiro}}]{2006MNRAS.372..679M}
{Mellema} G., {Iliev} I.~T., {Pen} U.-L., {Shapiro} P.~R., 2006{\natexlab{b}},
  \mnras, 372, 679

\bibitem[{{Mellema} {et~al}\mbox{.}(2013){Mellema}, {Koopmans}, {Abdalla},
  {Bernardi}, {Ciardi}, {Daiboo}, {de Bruyn}, {Datta}, {Falcke}, {Ferrara},
  {Iliev}, {Iocco}, {Jeli{\'c}}, {Jensen}, {Joseph}, {Labroupoulos}, {Meiksin},
  {Mesinger}, {Offringa}, {Pandey}, {Pritchard}, {Santos}, {Schwarz},
  {Semelin}, {Vedantham}, {Yatawatta}, \& {Zaroubi}}]{2013ExA....36..235M}
{Mellema} G. {et~al.}, 2013, Experimental Astronomy, 36, 235

\bibitem[{{Merz}, {Pen} \& {Trac}(2005){Merz}, {Pen}, \&
  {Trac}}]{2005NewA...10..393M}
{Merz} H., {Pen} U.-L., {Trac} H., 2005, \na, 10, 393

\bibitem[{{Mesinger} \& {Furlanetto}(2007)}]{2007ApJ...669..663M}
{Mesinger} A., {Furlanetto} S., 2007, \apj, 669, 663

\bibitem[{{Morales} \& {Hewitt}(2004)}]{2004ApJ...615....7M}
{Morales} M.~F., {Hewitt} J., 2004, \apj, 615, 7

\bibitem[{{Morales} \& {Wyithe}(2010)}]{2010ARA&A..48..127M}
{Morales} M.~F., {Wyithe} J.~S.~B., 2010, \araa, 48, 127

\bibitem[{{Nusser}(2005)}]{2005MNRAS.364..743N}
{Nusser} A., 2005, \mnras, 364, 743

\bibitem[{{Paciga} {et~al}\mbox{.}(2013){Paciga}, {Albert}, {Bandura}, {Chang},
  {Gupta}, {Hirata}, {Odegova}, {Pen}, {Peterson}, {Roy}, {Shaw}, {Sigurdson},
  \& {Voytek}}]{2013MNRAS.433..639P}
{Paciga} G. {et~al.}, 2013, \mnras, 433, 639

\bibitem[{{Parsons} {et~al}\mbox{.}(2010){Parsons}, {Backer}, {Foster},
  {Wright}, {Bradley}, {Gugliucci}, {Parashare}, {Benoit}, {Aguirre}, {Jacobs},
  {Carilli}, {Herne}, {Lynch}, {Manley}, \& {Werthimer}}]{2010AJ....139.1468P}
{Parsons} A.~R. {et~al.}, 2010, \aj, 139, 1468

\bibitem[{{Parsons} {et~al}\mbox{.}(2013){Parsons}, {Liu}, {Aguirre}, {Ali},
  {Bradley}, {Carilli}, {DeBoer}, {Dexter}, {Gugliucci}, {Jacobs}, {Klima},
  {MacMahon}, {Manley}, {Moore}, {Pober}, {Stefan}, \&
  {Walbrugh}}]{2013arXiv1304.4991P}
{Parsons} A.~R. {et~al.}, 2013, arXiv-1304.4991

\bibitem[{{Patil} {et~al}\mbox{.}(2014){Patil}, {Zaroubi}, {Chapman},
  {Jeli{\'c}}, {Harker}, {Abdalla}, {Asad}, {Bernardi}, {Brentjens}, {de
  Bruyn}, {Bus}, {Ciardi}, {Daiboo}, {Fernandez}, {Ghosh}, {Jensen}, {Kazemi},
  {Koopmans}, {Labropoulos}, {Mevius}, {Martinez}, {Mellema}, {Offringa},
  {Pandey}, {Schaye}, {Thomas}, {Vedantham}, {Veligatla}, {Wijnholds}, \&
  {Yatawatta}}]{2014arXiv1401.4172P}
{Patil} A.~H. {et~al.}, 2014, arXiv-1401.4172

\bibitem[{{Peebles}(1980)}]{1980lssu.book.....P}
{Peebles} P.~J.~E., 1980, {The large-scale structure of the universe}.
  Princeton University Press, 1980.~435 p.

\bibitem[{{Pen} {et~al}\mbox{.}(2009){Pen}, {Chang}, {Hirata}, {Peterson},
  {Roy}, {Gupta}, {Odegova}, \& {Sigurdson}}]{2009MNRAS.399..181P}
{Pen} U.-L., {Chang} T.-C., {Hirata} C.~M., {Peterson} J.~B., {Roy} J., {Gupta}
  Y., {Odegova} J., {Sigurdson} K., 2009, \mnras, 399, 181

\bibitem[{{Pen} {et~al}\mbox{.}(2008){Pen}, {Chang}, {Peterson}, {Roy},
  {Gupta}, \& {Bandura}}]{2008AIPC.1035...75P}
{Pen} U.-L., {Chang} T.-C., {Peterson} J.~B., {Roy} J., {Gupta} Y., {Bandura}
  K., 2008, in American Institute of Physics Conference Series, Vol. 1035, The
  Evolution of Galaxies Through the Neutral Hydrogen Window, {Minchin} R.,
  {Momjian} E., eds., pp. 75--81

\bibitem[{{Planck Collaboration} {et~al}\mbox{.}(2013){Planck Collaboration},
  {Ade}, {Aghanim}, {Armitage-Caplan}, {Arnaud}, {Ashdown}, {Atrio-Barandela},
  {Aumont}, {Baccigalupi}, {Banday}, \& et~al.}]{2013arXiv1303.5076P}
{Planck Collaboration} {et~al.}, 2013, ArXiv e-prints

\bibitem[{{Pritchard} \& {Loeb}(2012)}]{2012RPPh...75h6901P}
{Pritchard} J.~R., {Loeb} A., 2012, Reports on Progress in Physics, 75, 086901

\bibitem[{{Santos} {et~al}\mbox{.}(2008){Santos}, {Amblard}, {Pritchard},
  {Trac}, {Cen}, \& {Cooray}}]{2008ApJ...689....1S}
{Santos} M.~G., {Amblard} A., {Pritchard} J., {Trac} H., {Cen} R., {Cooray} A.,
  2008, \apj, 689, 1

\bibitem[{{Sethi} \& {Haiman}(2008)}]{2008ApJ...673....1S}
{Sethi} S., {Haiman} Z., 2008, \apj, 673, 1

\bibitem[{{Shapiro} {et~al}\mbox{.}(2006){Shapiro}, {Iliev}, {Alvarez}, \&
  {Scannapieco}}]{2006ApJ...648..922S}
{Shapiro} P.~R., {Iliev} I.~T., {Alvarez} M.~A., {Scannapieco} E., 2006, \apj,
  648, 922

\bibitem[{{Shapiro} {et~al}\mbox{.}(2013){Shapiro}, {Mao}, {Iliev}, {Mellema},
  {Datta}, {Ahn}, \& {Koda}}]{2013PhRvL.110o1301S}
{Shapiro} P.~R., {Mao} Y., {Iliev} I.~T., {Mellema} G., {Datta} K.~K., {Ahn}
  K., {Koda} J., 2013, Physical Review Letters, 110, 151301

\bibitem[{{Shin}, {Trac} \& {Cen}(2008){Shin}, {Trac}, \&
  {Cen}}]{2008ApJ...681..756S}
{Shin} M.-S., {Trac} H., {Cen} R., 2008, \apj, 681, 756

\bibitem[{{Thomas} {et~al}\mbox{.}(2009){Thomas}, {Zaroubi}, {Ciardi},
  {Pawlik}, {Labropoulos}, {Jeli{\'c}}, {Bernardi}, {Brentjens}, {de Bruyn},
  {Harker}, {Koopmans}, {Mellema}, {Pandey}, {Schaye}, \&
  {Yatawatta}}]{2009MNRAS.393...32T}
{Thomas} R.~M. {et~al.}, 2009, \mnras, 393, 32

\bibitem[{{Tingay} {et~al}\mbox{.}(2013){Tingay}, {Goeke}, {Bowman}, {Emrich},
  {Ord}, {Mitchell}, {Morales}, {Booler}, {Crosse}, {Wayth}, {Lonsdale},
  {Tremblay}, {Pallot}, {Colegate}, {Wicenec}, {Kudryavtseva}, {Arcus},
  {Barnes}, {Bernardi}, {Briggs}, {Burns}, {Bunton}, {Cappallo}, {Corey},
  {Deshpande}, {Desouza}, {Gaensler}, {Greenhill}, {Hall}, {Hazelton}, {Herne},
  {Hewitt}, {Johnston-Hollitt}, {Kaplan}, {Kasper}, {Kincaid}, {Koenig},
  {Kratzenberg}, {Lynch}, {Mckinley}, {Mcwhirter}, {Morgan}, {Oberoi},
  {Pathikulangara}, {Prabu}, {Remillard}, {Rogers}, {Roshi}, {Salah}, {Sault},
  {Udaya-Shankar}, {Schlagenhaufer}, {Srivani}, {Stevens}, {Subrahmanyan},
  {Waterson}, {Webster}, {Whitney}, {Williams}, {Williams}, \&
  {Wyithe}}]{2013PASA...30....7T}
{Tingay} S.~J. {et~al.}, 2013, Publications of the Astronomical Society of
  Australia (PASA), 30, 7

\bibitem[{{van Haarlem} {et~al}\mbox{.}(2013){van Haarlem}, {Wise}, {Gunst},
  {Heald}, {McKean}, {Hessels}, {de Bruyn}, {Nijboer}, {Swinbank}, {Fallows},
  {Brentjens}, {Nelles}, {Beck}, {Falcke}, {Fender}, {H{\"o}randel},
  {Koopmans}, {Mann}, {Miley}, {R{\"o}ttgering}, {Stappers}, {Wijers},
  {Zaroubi}, {van den Akker}, {Alexov}, {Anderson}, {Anderson}, {van Ardenne},
  {Arts}, {Asgekar}, {Avruch}, {Batejat}, {B{\"a}hren}, {Bell}, {Bell}, {van
  Bemmel}, {Bennema}, {Bentum}, {Bernardi}, {Best}, {B{\^i}rzan}, {Bonafede},
  {Boonstra}, {Braun}, {Bregman}, {Breitling}, {van de Brink}, {Broderick},
  {Broekema}, {Brouw}, {Br{\"u}ggen}, {Butcher}, {van Cappellen}, {Ciardi},
  {Coenen}, {Conway}, {Coolen}, {Corstanje}, {Damstra}, {Davies}, {Deller},
  {Dettmar}, {van Diepen}, {Dijkstra}, {Donker}, {Doorduin}, {Dromer}, {Drost},
  {van Duin}, {Eisl{\"o}ffel}, {van Enst}, {Ferrari}, {Frieswijk}, {Gankema},
  {Garrett}, {de Gasperin}, {Gerbers}, {de Geus}, {Grie{\ss}meier}, {Grit},
  {Gruppen}, {Hamaker}, {Hassall}, {Hoeft}, {Holties}, {Horneffer}, {van der
  Horst}, {van Houwelingen}, {Huijgen}, {Iacobelli}, {Intema}, {Jackson},
  {Jelic}, {de Jong}, {Juette}, {Kant}, {Karastergiou}, {Koers}, {Kollen},
  {Kondratiev}, {Kooistra}, {Koopman}, {Koster}, {Kuniyoshi}, {Kramer},
  {Kuper}, {Lambropoulos}, {Law}, {van Leeuwen}, {Lemaitre}, {Loose}, {Maat},
  {Macario}, {Markoff}, {Masters}, {McFadden}, {McKay-Bukowski}, {Meijering},
  {Meulman}, {Mevius}, {Middelberg}, {Millenaar}, {Miller-Jones}, {Mohan},
  {Mol}, {Morawietz}, {Morganti}, {Mulcahy}, {Mulder}, {Munk}, {Nieuwenhuis},
  {van Nieuwpoort}, {Noordam}, {Norden}, {Noutsos}, {Offringa}, {Olofsson},
  {Omar}, {Orr{\'u}}, {Overeem}, {Paas}, {Pandey-Pommier}, {Pandey}, {Pizzo},
  {Polatidis}, {Rafferty}, {Rawlings}, {Reich}, {de Reijer}, {Reitsma},
  {Renting}, {Riemers}, {Rol}, {Romein}, {Roosjen}, {Ruiter}, {Scaife}, {van
  der Schaaf}, {Scheers}, {Schellart}, {Schoenmakers}, {Schoonderbeek},
  {Serylak}, {Shulevski}, {Sluman}, {Smirnov}, {Sobey}, {Spreeuw}, {Steinmetz},
  {Sterks}, {Stiepel}, {Stuurwold}, {Tagger}, {Tang}, {Tasse}, {Thomas},
  {Thoudam}, {Toribio}, {van der Tol}, {Usov}, {van Veelen}, {van der Veen},
  {ter Veen}, {Verbiest}, {Vermeulen}, {Vermaas}, {Vocks}, {Vogt}, {de Vos},
  {van der Wal}, {van Weeren}, {Weggemans}, {Weltevrede}, {White}, {Wijnholds},
  {Wilhelmsson}, {Wucknitz}, {Yatawatta}, {Zarka}, {Zensus}, \& {van
  Zwieten}}]{2013A&A...556A...2V}
{van Haarlem} M.~P. {et~al.}, 2013, \aap, 556, A2

\bibitem[{{Wyithe}, {Loeb} \& {Barnes}(2005){Wyithe}, {Loeb}, \&
  {Barnes}}]{2005ApJ...634..715W}
{Wyithe} J.~S.~B., {Loeb} A., {Barnes} D.~G., 2005, \apj, 634, 715

\bibitem[{{Yatawatta} {et~al}\mbox{.}(2013){Yatawatta}, {de Bruyn},
  {Brentjens}, {Labropoulos}, {Pandey}, {Kazemi}, {Zaroubi}, {Koopmans},
  {Offringa}, {Jeli{\'c}}, {Martinez Rubi}, {Veligatla}, {Wijnholds}, {Brouw},
  {Bernardi}, {Ciardi}, {Daiboo}, {Harker}, {Mellema}, {Schaye}, {Thomas},
  {Vedantham}, {Chapman}, {Abdalla}, {Alexov}, {Anderson}, {Avruch}, {Batejat},
  {Bell}, {Bell}, {Bentum}, {Best}, {Bonafede}, {Bregman}, {Breitling}, {van de
  Brink}, {Broderick}, {Br{\"u}ggen}, {Conway}, {de Gasperin}, {de Geus},
  {Duscha}, {Falcke}, {Fallows}, {Ferrari}, {Frieswijk}, {Garrett},
  {Griessmeier}, {Gunst}, {Hassall}, {Hessels}, {Hoeft}, {Iacobelli}, {Juette},
  {Karastergiou}, {Kondratiev}, {Kramer}, {Kuniyoshi}, {Kuper}, {van Leeuwen},
  {Maat}, {Mann}, {McKean}, {Mevius}, {Mol}, {Munk}, {Nijboer}, {Noordam},
  {Norden}, {Orru}, {Paas}, {Pandey-Pommier}, {Pizzo}, {Polatidis}, {Reich},
  {R{\"o}ttgering}, {Sluman}, {Smirnov}, {Stappers}, {Steinmetz}, {Tagger},
  {Tang}, {Tasse}, {ter Veen}, {Vermeulen}, {van Weeren}, {Wise}, {Wucknitz},
  \& {Zarka}}]{2013A&A...550A.136Y}
{Yatawatta} S. {et~al.}, 2013, \aap, 550, A136

\bibitem[{{Yu}(2005)}]{2005ApJ...623..683Y}
{Yu} Q., 2005, \apj, 623, 683

\bibitem[{{Zahn} {et~al}\mbox{.}(2007){Zahn}, {Lidz}, {McQuinn}, {Dutta},
  {Hernquist}, {Zaldarriaga}, \& {Furlanetto}}]{2007ApJ...654...12Z}
{Zahn} O., {Lidz} A., {McQuinn} M., {Dutta} S., {Hernquist} L., {Zaldarriaga}
  M., {Furlanetto} S.~R., 2007, \apj, 654, 12

\bibitem[{{Zaldarriaga}, {Furlanetto} \& {Hernquist}(2004){Zaldarriaga},
  {Furlanetto}, \& {Hernquist}}]{2004ApJ...608..622Z}
{Zaldarriaga} M., {Furlanetto} S.~R., {Hernquist} L., 2004, \apj, 608, 622

\bibitem[{{Zaroubi} {et~al}\mbox{.}(2012){Zaroubi}, {de Bruyn}, {Harker},
  {Thomas}, {Labropolous}, {Jeli{\'c}}, {Koopmans}, {Brentjens}, {Bernardi},
  {Ciardi}, {Daiboo}, {Kazemi}, {Martinez-Rubi}, {Mellema}, {Offringa},
  {Pandey}, {Schaye}, {Veligatla}, {Vedantham}, \&
  {Yatawatta}}]{2012MNRAS.425.2964Z}
{Zaroubi} S. {et~al.}, 2012, \mnras, 425, 2964

\bibitem[{{Zawada} {et~al}\mbox{.}(2014){Zawada}, {Semelin}, {Vonlanthen},
  {Baek}, \& {Revaz}}]{2014arXiv1401.1807Z}
{Zawada} K., {Semelin} B., {Vonlanthen} P., {Baek} S., {Revaz} Y., 2014,
  MNRAS(in press), arXiv-1401.1807

\end{thebibliography}
\bibliographystyle{mn2e}

\label{lastpage}
\end{document}